\documentclass[a4paper,11pt]{article}
\pdfoutput=1 % if your are submitting a pdflatex (i.e. if you have
\usepackage{jcappub} % for details on the use of the package, please
\usepackage[T1]{fontenc} % if needed

\usepackage{mathalfa}
\usepackage{graphicx}
\usepackage{caption}
\usepackage{subcaption}
\usepackage{amsmath,amssymb}
\usepackage[font=small,labelfont=bf]{caption}
\usepackage{verbatim}
\usepackage{placeins}

\def\mean#1{\left< #1 \right>}

\usepackage{color}

\usepackage[normalem]{ulem}

\title{The Alcock Paczy\'nski  test with Baryon Acoustic Oscillations: systematic effects for future surveys}

\newcommand{\bx}{{\mathbf x}}

\newcommand{\bn}{{\mathbf n}}

\newcommand{\De}{\Delta}

\newcommand{\be}{\begin{equation}}
\newcommand{\ee}{\end{equation}}

\newcommand{\lsim}{\stackrel{<}{\sim}}
\newcommand{\bea}{\begin{eqnarray}}
\newcommand{\eea}{\end{eqnarray}}
\newcommand{\bean}{\begin{eqnarray*}}
\newcommand{\eean}{\end{eqnarray*}}

\newcommand{\HH}{{\cal H}}
\author[a, b]{Francesca Lepori,}
\author[c, b]{Enea Di Dio,}
\author[a, c, b]{Matteo Viel,}
\author[a, b, c]{\\Carlo Baccigalupi,}
\author[d]{Ruth Durrer}

\affiliation[a]{SISSA- International School for Advanced Studies, \\ Via Bonomea 265, 34136 Trieste, Italy}
\affiliation[b]{INFN, Sezione di Trieste,\\ Via Valerio 2, I-34127 Trieste, Italy}
\affiliation[c]{INAF - Osservatorio Astronomico di Trieste,\\  Via G. B. Tiepolo 11,  I-34143 Trieste, Italy}
\affiliation[d]{Universit\'{e}  de  Gen\`{e}ve,  D\'{e}partement  de  Physique  Th\'{e}orique  and  CAP,\\  24  quai  Ernest-Ansermet, CH-1211 Gen\`{e}ve 4, Switzerland}

\emailAdd{flepori@sissa.it}
\emailAdd{enea.didio@oats.inaf.it}
\emailAdd{viel@oats.inaf.it}
\emailAdd{carlo.baccigalupi@sissa.it}
\emailAdd{Ruth.Durrer@unige.ch}

\abstract{
We investigate the Alcock Paczy\'nski  (AP) test applied to the Baryon Acoustic Oscillation (BAO) feature in the galaxy correlation function. By using a general formalism that includes relativistic effects, we  quantify the importance of the linear redshift space distortions and gravitational lensing corrections to the galaxy number density fluctuation.
We show that redshift space distortions significantly affect the shape of the correlation function, both 
in radial and transverse directions, causing different values of galaxy bias to induce offsets up to 1\% in the AP test.
On the other hand, we find that the lensing correction around the BAO scale modifies the amplitude but not the shape of the correlation function and therefore does not introduce any systematic effect.
Furthermore, we investigate in details how the AP test is sensitive to redshift binning:
a window function in transverse direction  suppresses  correlations
and shifts the peak position toward smaller angular scales. 
We determine the correction that should be applied in order to account for this effect,
when performing the test with data from three future planned galaxy redshift surveys: Euclid, the Dark Energy Spectroscopic Instrument (DESI) and the Square Kilometer Array (SKA).
}

\begin{document}
\maketitle
\flushbottom
\section{Introduction}

With the measurements of Cosmic Microwave Background (CMB) anisotropies, performed by WMAP~\cite{wmap} and Planck~\cite{planck2015} experiments, cosmology has entered a precision era, leading to the confirmation of $\Lambda$CDM as the standard model for cosmology. 
Despite the stunning agreement between theory and data that has been reached, still
many open questions remain to be investigated within the standard model.
One of these unsolved puzzles is the accelerated expansion of the universe~\cite{DE}, which
in the standard cosmological model is driven by a cosmological constant $\Lambda$, whose value leads to a serious fine-tuning problem and a coincidence problem \cite{ccproblem, Frieman:2008sn}. Therefore many alternative models with dynamical dark energy or modifications of General Relativity which can lead to the same accelerated expansion as $\Lambda$CDM have been proposed~\cite{Durrer:2007re}.
A crucial role in shedding light on the nature and origin of cosmic acceleration will be played 
in the next decades by Large Scale Structure (LSS) experiments, which will
quantify the impact of Dark Energy in the growth of cosmic structures.
Different tracers of cosmic structures will provide information on different stages of the expansion history.

In this work we focus on galaxy clustering. The upcoming galaxy surveys will
map the distribution of galaxies on a large fraction of the sky up to redshift $z \sim 2$.
In order to exploit this huge  amount of incoming data, an accurate model for 
what we will be measuring in galaxy surveys is required. 
Moreover, having  much better statistics, we may release some of the  assumptions and proceed in a more model independent way.

In the past few years, the galaxy number counts have been computed including all the relativistic effects at the linear order~\cite{yooA,yooB,bonvin_durrer, challinor_lewis} and at second order~\cite{Yoo:2014sfa,Bertacca:2014dra,DiDio:2014lka} in perturbation theory. Besides the known redshift space distortions correction to the local overdensity of
galaxies~\cite{kaiser}, other terms contribute to galaxy number counts,  e.g.~Doppler corrections 
and gravitational lensing. Not taking into account these effects in our theoretical model may bias the analysis~\cite{Cardona:2016qxn}, therefore their relevance in any cosmological observable should be tested.

On relatively small scales, other effects such as the non-linear redshift
space distortions (the so called "finger of god" effect) may be relevant. In this work we do not model these distortions. We focus on the regime of large scales, where linear theory
is assumed to be a good approximation.

In this paper we will investigate the relevance of these corrections for the Alcock Paczy\'nski (AP) test~\cite{AP_original}.
We start from the method proposed in~\cite{montanari_durrer}, where the AP test
is performed on the Baryon Acoustic Oscillation (BAO) feature of the galaxy 2-point correlation function without any prior assumption on the cosmological parameters. Nevertheless, we will show that some prior information about the galaxy bias can improve the accuracy in determining the BAO scale from the observable quantities.

In Section~\ref{GR_eff} we introduce the formalism we employ throughout the paper for the observed galaxy number density.
In Section~\ref{sec3} we present the AP test and we introduce the AP parameter.
In Section~\ref{sec4} we summarize the method we use to compute the galaxy 2-point correlation function and to determine the position of the acoustic peak and we outline the general strategy to
investigate observational distortions of the test. More details on the methodology are given in Appendix~\ref{Ap:B}.
In Section~\ref{sec5} we report our results: in~\ref{sec5a}, we discuss the relevance of the relativistic correction to the galaxy number count for the AP test, in~\ref{sec5b} we introduce a linear local bias and we show how it affects our method;
in~\ref{window} we study projection effects induced by a radial window function and 
we compute the corrective factors that must be applied to the estimated AP parameter for three future planned galaxy surveys, i.e.~Euclid~\cite{Amendola}, the Dark Energy Spectroscopic Instrument (DESI)~\cite{desi0}
and the Square Kilometer Array (SKA)~\cite{Abdalla}. We also analyze the impact of shot-noise and cosmic variance on the accuracy of the BAO peak determination in Section~\ref{sec:shotnoise}.
In Section~\ref{conc} we draw the conclusions.

\section{Relativistic formalism for galaxy correlations}
\label{GR_eff}
We consider a Friedmann universe with linear scalar perturbations. We work in Newtonian gauge so that the line element is given by
\be
ds^2 = a(t)^2 \left( -\left( 1+ 2 \Psi \right) dt^2 + \left( 1 - 2 \Phi \right) d\bx^2 \right) \, ,
\ee
where $a(t)$ is the scale factor, $t$ is conformal time and the metric perturbations, $\Psi$ and $\Phi$, are the Bardeen potentials. We remark that working solely with observational quantities, we can fix the gauge without loss of generality.

Galaxies are discrete tracers of the density field of the universe. In order to obtain
cosmological information from the analysis of their distribution, it is
crucial to identify the true observable in galaxy surveys and how the large scale dark matter distribution, that we cannot directly observe, is mapped into observable coordinates.
The true observable in galaxy surveys, the galaxy number density fluctuation $\Delta_\text{obs}$, has been computed at first order
in perturbation theory using a relativistic formalism by tracing
the photons path between the source and the observer~\cite{yooA,yooB, bonvin_durrer, challinor_lewis}.
These are expressed as functions of the observed redshift of the source $z$ and the direction of propagation of the emitted photons $\bn$. It includes a local density field term plus additional terms due to fact that measured redshift $z$ and angular position\footnote{We refer to the angular position with $\bn$, where $\bn$ denotes a unit vector (direction) in the celestial sphere.} ($\bn$) are perturbed by velocity and metric perturbations at the source position and by the intervening gravitational potentials between the source and the observer.

The total observed galaxy overdensity $\Delta_\text{obs} \left( \bn , z \right) $ can be schematically expressed as a sum of different contributions

\begin{equation}
\label{1}
\Delta_\text{obs} = \Delta_g + \Delta_\text{RSD} + \Delta_{\kappa} + \Delta_\text{rel},
\end{equation}
where $\Delta_g$ is the local galaxy overdensity,
\begin{equation}
  \Delta_g= b  \  \delta
\end{equation}
proportional to the density contrast in comoving gauge  $\delta$. 
For the sake of simplicity, we assume the bias $b$ to be linear
and local.
The term $\Delta_\text{RSD}$ in our notation includes the linear 
redshift space distortions contributions, due to the peculiar motion of galaxies, plus other subdominant Doppler corrections~\cite{Enea}
\begin{equation}
\Delta_\text{RSD}(\mathbf{n}, z) = \frac{1}{\mathcal{H}(z)} \partial_r (\mathbf{V}\cdot\mathbf{n}) + \Biggl(\frac{\mathcal{H'}}{\mathcal{H}^2} + \frac{2}{r\mathcal{H}}\Biggr)(\mathbf{V}\cdot \mathbf{n})
 - 3\HH V , \label{RSDcorr}
\end{equation}
where $\mathbf{V}$ is the peculiar velocity in longitudinal gauge, $V$ the potential velocity defined by ${\bf V} = -{\bf \nabla} V$, $r = t_o -t$ is the conformal distance
and $\mathcal{H}=a'/a$ is the conformal Hubble factor and a prime denotes a derivative with respect to the conformal time $t$.
$\Delta_{\kappa}$ is the gravitational lensing term,
\begin{equation}
\Delta_{\kappa} = -\int_0^{r(z)} \frac{r(z) - r}{r(z) r} \Delta_{\Omega} (\Phi + \Psi) dr,
\end{equation}
where $\Delta_{\Omega}$
is the Laplace operator on the sphere.
The last term in Eq.~\eqref{1}, $ \Delta_{\rm rel},$ includes subdominant
local and integrated combinations of the Bardeen potentials:
\begin{align}
\Delta_{\rm rel} = &- 2 \Phi + \Psi + \frac{1}{\mathcal{H}} \Phi'
              +\frac{2}{r(z)} \int^{r(z)}_0 dr (\Phi + \Psi) + \nonumber  \\
              & +\Biggl(\frac{\mathcal{H'}}{\mathcal{H}^2} + \frac{2}{r(z)\mathcal{H}}\Biggr)\Biggl(\Psi + \int^{r(z)}_0 dr (\Phi' + \Psi')\Biggr)\, . \label{Delta_rel}
\end{align}
Throughout this work we will neglect the effect of magnification bias and the possible evolution of the number of counts~\cite{challinor_lewis}. Even if present,  these bias factors multiply terms which are subdominant in the AP test.

\begin{figure}
    \begin{subfigure}[b]{0.5\textwidth}
        \includegraphics[width=\textwidth]{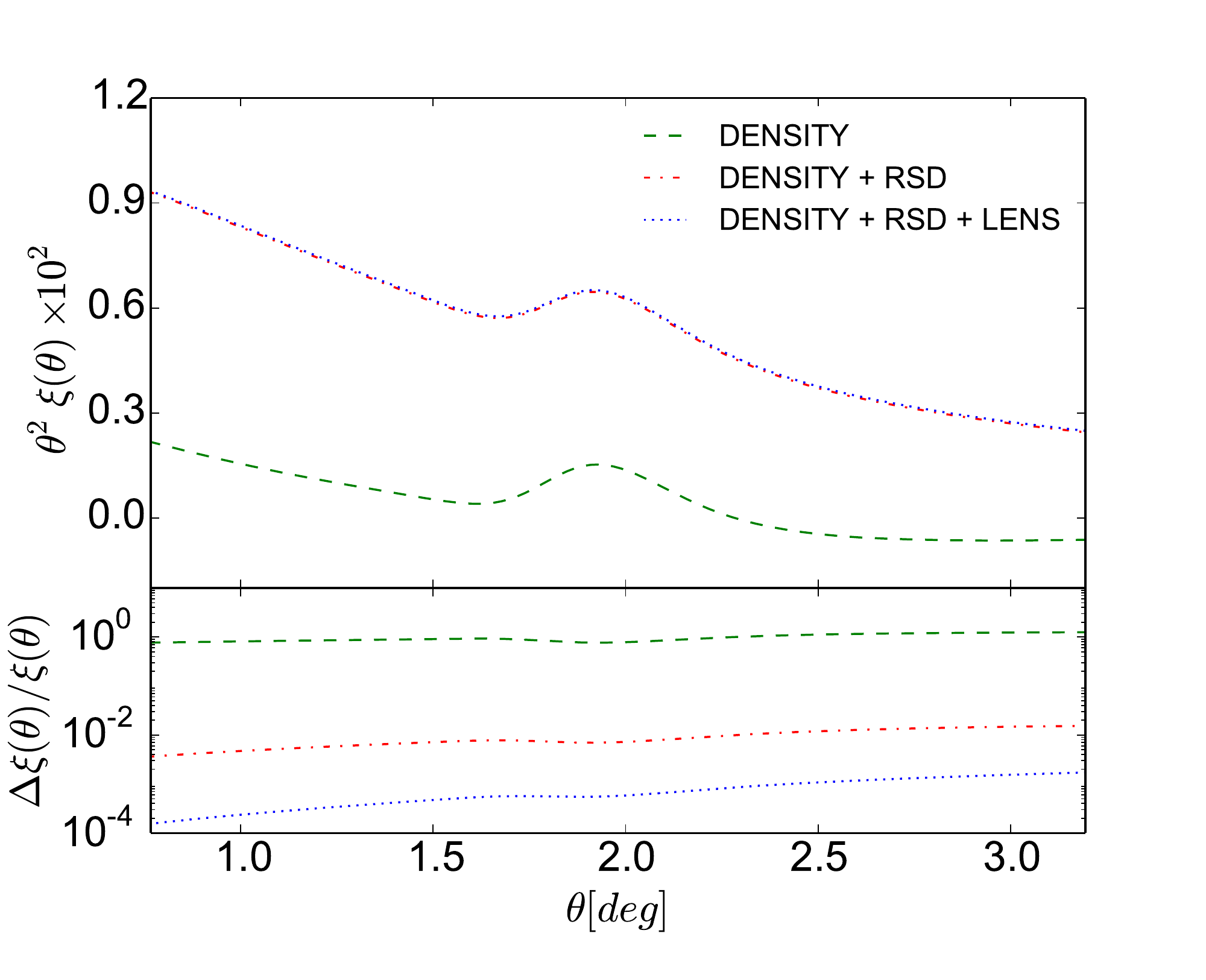}
        \caption{Transverse correlation function}
        \label{fig:Transv}
    \end{subfigure}
    \begin{subfigure}[b]{0.5\textwidth}
        \includegraphics[width=\textwidth]{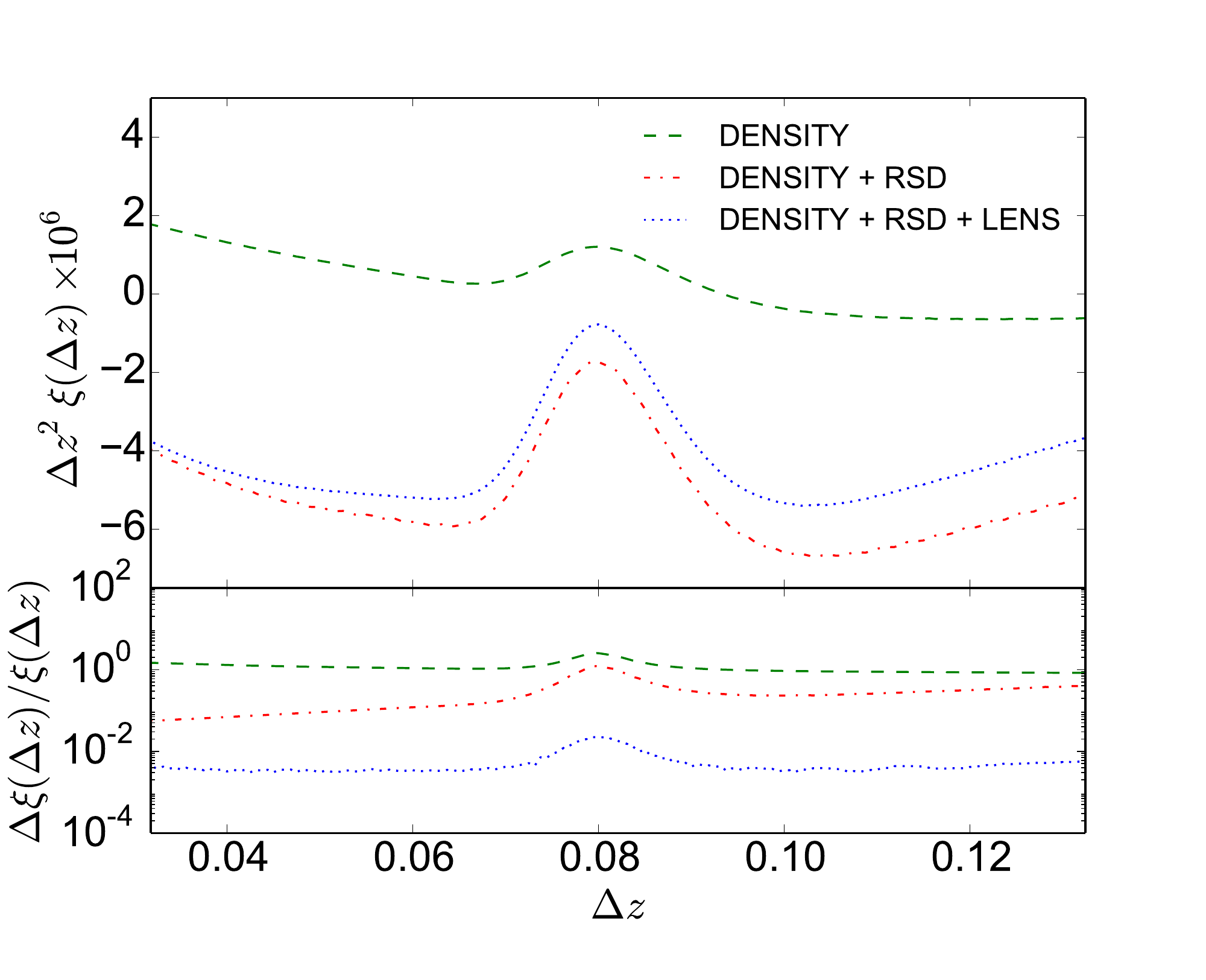}
        \caption{Longitudinal correlation function}
        \label{fig:Long}
    \end{subfigure}
    \caption{Transverse and longitudinal correlation function computed at $z_{\rm mean} = 1.5$, partially including the corrective terms in Eq.~\eqref{1}. 
In the bottom panels, we show the relative differences between the exact correlation function (including all the terms) and a partial correlation function computed including only the local density term (green, dashed line), density and redshift space distortion
correction (red, dash-dotted line),
the previous terms plus the lensing (blue, dotted line).} \label{fig:GR_eff}
\end{figure}

In terms of observational coordinates, the relevant statistical quantities are
the angular correlation function $\xi (\theta, z_1, z_2)$ or the redshift dependent angular power spectra $C_{\ell} (z_1, z_2)$.
The galaxy correlation function, under the assumption of statistical isotropy, reduces to
\begin{equation}
\xi (\theta, z_1, z_2) = \mean{\Delta_\text{obs}(\mathbf{n}_1, z_1)\Delta_\text{obs}(\mathbf{n}_2, z_2)}, \qquad \cos{\theta} \equiv \mathbf{n_1}
\cdot \mathbf{n_2},
\end{equation}
where $\langle .. \rangle$ denotes the ensemble average over several realizations. Observationally, this is replaced by an average over directions at fixed observed redshift  and opening angle $\theta$. 
In a similar way, the angular power spectrum is defined as~\cite{bonvin_durrer}
\begin{equation}
C_{\ell} (z_1, z_2) = \mean{a_{\ell m}(z_1)a^*_{ \ell m}(z_2)},
\end{equation}
where a star denotes the complex conjugate and $a_{\ell m}$ are the coefficients of the spherical harmonic
expansion for  $\Delta(\mathbf{n}, z)$
\begin{equation}
\Delta_\text{obs}(\mathbf{n}, z) = \sum_{\ell m} a_{\ell m}(z) Y_{ \ell m}(\mathbf{n}),
\qquad a_{\ell m} \left( z \right)  = \int d \Omega_{\mathbf{n}} Y^*_{\ell m} (\mathbf{n})
\Delta_\text{obs}(\mathbf{n}, z).
\end{equation}
The angular power spectra and the correlation functions are related by
\begin{equation}
\xi(\theta, z_1, z_2) = \frac{1}{4\pi}\sum_{\ell=0}^{\infty} (2\ell+1)
C_{\ell} (z_1, z_2) P_{\ell}(\cos{\theta}), \label{xi_from_cl}
\end{equation}
where $P_{\ell}(\cos{\theta})$ is the Legendre polynomial of degree $\ell$.
From Eq.~\eqref{xi_from_cl} it is straightforward to define the correlation
function along the line-of-sight direction by setting $\theta=0$ and in the transverse direction by setting $z_1=z_2$.
 
The correlation function at a given mean redshift $z_\text{mean}$ along the line-of-sight is given by 
\begin{equation}
\xi_{\parallel}(\Delta z, z_\text{mean}) = \frac{1}{4\pi}\sum_{\ell=0}^{\infty} (2\ell+1)
C_{\ell} (z_\text{mean} - \Delta z/2, z_\text{mean} + \Delta z/2) 
\label{Long_corr}
\end{equation}
while the transverse correlations are 
\begin{equation}
\xi_{\perp}(\theta, z_\text{mean}) = \frac{1}{4\pi}\sum_{\ell=0}^{\infty} (2\ell+1)
C_{\ell} (z_\text{mean}, z_\text{mean}) P_{\ell}(\cos{\theta}). \label{Transv_corr}
\end{equation}
Here $ z_\text{mean}=(z_1+z_2)/2$ and $\Delta z=z_2-z_1$.

Figure~\ref{fig:GR_eff} represents transverse and longitudinal correlation
function at fixed mean redshift $z_\text{mean} = 1.5$.
In each plot the correlation function is computed gradually adding
subdominant contributions.
Redshift-space distortions affect considerably the correlations
in both directions, while lensing does not change sensitively the transverse
correlations. In the longitudinal direction,  the lensing term enhances the amplitude of the correlation function by up to 30\% for $\Delta z\lsim 0.13$, but it does not modify its shape at the BAO scale. We notice that the lensing effect is more important for pairs of galaxies with large radial separation. The contribution from the relativistic terms, i.e.~Eq.~(\ref{Delta_rel}), is completely subdominant and it would not be visible in Figure~\ref{fig:GR_eff}. Therefore we neglect it in the analysis performed in the rest of the paper. We remark that the largest relativistic correction, namely the Doppler term, is included in the redshift space distortions, see Eq.~(\ref{RSDcorr}).

\section{The Alcock Paczy\'nski test}
\label{sec3}
The Alcock Paczy\'nski test~\cite{AP_original}, proposed for the first time in 1979, is a purely geometrical test of the
cosmic expansion history performed by measuring the shape of an object expanding with the Hubble flow. 
When we observe an astrophysical object, we measure its
shape in terms of its angular size $\theta$ and its radial extent in redshift space $\Delta z$. 
These two quantities depend on the comoving sizes of the source, $L_\parallel$ and $L_\perp$, and on a conversion factor: $\Delta z$ is related
to the Hubble expansion rate $H\left(z\right) $ by
\begin{equation}
\Delta z = L_\parallel H(z),
\end{equation}
 while the dependence of $\theta$ on the angular diameter distance
is given by
 \begin{equation}
 \theta = \frac{L_\perp}{(1+z)D_A(z)}.
 \end{equation}
If the object is known to be spherically symmetric, i.e.~$L=L_\perp= L_\parallel$, then 
the ratio of these two measured quantities,
 \begin{equation}
 \label{eq:Fap}
 F_{\rm AP} \equiv \frac{\Delta z}{\theta},
 \end{equation} 
 does not depend on the physical size of the object, but only on the redshift and on the spacetime geometry
\begin{equation}
 F_{\rm th}(z) = (1+z) D_A(z) H(z). 
 \end{equation}  
Here we explicitly distinguished the measured AP parameter $F_{\rm AP}$ from its theoretical (background) value $F_{\rm th}$.

In realistic applications, the spherical symmetry is not required
to hold for single objects because the test can be applied to the galaxy clustering statistics. 
In fact,  statistical isotropy of space implies statistical spherical symmetry for the correlation function.
Furthermore, the correlation function naturally offers a robust
feature for the application of the AP test, the BAO scale. BAO high-precision measurements are one of the main target of future
spectroscopic surveys (see~\cite{BAO_2} for a recent overview of the
cosmological implication of BAO measurements).

The application of the AP test on the BAO feature in the galaxy correlation function has been proposed in~\cite{montanari_durrer},
where it has been found that the peak position of the transverse correlation function is significantly affected by the binning in redshift.
This effect is due to the fact that a  radial window 
for the transverse correlation function
introduces a spurious radial component in the correlation function
and the BAO peak estimation must be properly corrected for.
In this paper we aim to extend the work presented in~\cite{montanari_durrer}, addressing also other effects that may distort the result of the test. 
 
\section{Methodology}
\label{sec4}
We assume a $\Lambda$CDM cosmology consistent 
with the best fit parameters from the Planck 2015 data~\cite{planck2015}: 
$h = 0.6774$, $\Omega_{cdm}h^2 = 0.1188$, $\Omega_b h^2 = 0.0223$, $\Omega_{\Lambda} = 1-\Omega_{m}$, $\Omega_k = 0$.
The primordial amplitude and spectral index are set to 
$A_s = 2.142 \times 10^{-9}$ and $n_s = 0.9667$.

The computation of the angular power spectrum for the observed galaxy
number counts is implemented in the publicly available Boltzmann code
 {\sc class}gal~\cite{CLASSgal}, a modified version of the Cosmic Linear Anisotropy Solving System ({\sc class}) code~\cite{class1, class2} optimized 
to compute accurately and efficiently the relativistic large scale observables to linear order~\cite{Enea}.
The angular power spectrum is computed by running {\sc class}gal 
for the fiducial cosmology.
Unless otherwise stated, we do not include any window function in the model and 
we do not include non-linearities.

The radial and transverse correlation functions are computed
from~\eqref{Long_corr} and~\eqref{Transv_corr} summing over
a finite number of multipoles $\ell\leq\ell_{max}$. 
In the radial direction, the value $\ell_{max}$ can be set to be
 large enough in order to avoid spurious numerical oscillations induced by a sharp cutoff in $\ell$-space.
In the transverse direction we set $\ell_{max} = 20000$ and 
we introduced a cutoff $W_\ell$ to
smooth numerical spurious oscillations, so that we have
\begin{equation}
\xi_{\perp}(\theta) = \frac{1}{4\pi}\sum_{\ell=0}^{\ell_{max}} (2\ell+1)
C_{\ell} (z_{mean}, z_{mean}) P_\ell(\cos{\theta}) W_{\ell}(\ell_s, \ell_x), 
\end{equation}
where
\begin{equation}
W_{\ell}(\ell_s, \ell_x) = \frac{1}{2} \Bigl(1-\tanh{\{(\ell-\ell_s+3\ell_x)/\ell_x\}}\Bigr).
\end{equation} 
The cutoff parameters $\ell_s$ and $\ell_x$ are set to be respectively
$\ell_s \approx \ell_{max}$ and $\ell_x = 2000$.
The estimation of the BAO peak is not affected by
small variation of these parameters.
We remark that the correlation function can be computed directly for density, redshift space distortions and for local terms in general, see e.g. Ref.~\cite{Matsubara:1999du,Matsubara:2004fr,montanari_durrer}. Nevertheless we prefer to use Eq.~\eqref{Long_corr} and~\eqref{Transv_corr} to handle the integrated terms. In this way we can use the precise results obtained with {\sc class}gal code, instead of relying on some uncontrolled approximation.

We model the correlation functions with the following parameterization
\begin{equation}
\xi(x) = A \cdot e^{-(x-x_{\rm BAO})^2/2\sigma^2} + \sum^N_{n=0} K_n \cdot
x^n, \label{paramet}
\end{equation} 
where $x=\theta$ for the transverse correlation function and
$x = \Delta z$ for the radial one.
In Eq.~\eqref{paramet} a polynomial of degree $N$ models the shape of the correlation function on scales unaffected by the BAO peak,
while a Gaussian describes the BAO feature.

We fit the data points with the template model, where the
free parameters are $A$, $x_{\rm BAO}$, $\sigma$, $K_n$, with $n = 0,1,..., N$.
The BAO scale is estimated as the best fitting value of $\theta_{\rm BAO}$ and $\Delta z_{\rm BAO}$.

The non-linear least squares fitting is performed using the Python version of MPFIT~\cite{mpfit}, which implements the Levenberg-Marquardt method~\cite{LM}.

Once the position of the BAO feature has been estimated in
both radial and transverse directions,
we compute the AP parameter in Eq.~(\ref{eq:Fap}) 
and we compare it with its theoretical value $F_{\rm th}$, computed for
the same fiducial cosmology.
We perform the same analysis for different values of the 
mean redshift $z_\text{\rm mean}$ in the range between $z=0.3$ and $z=2$.
The two values are always expected to coincide, within error-bars.
Violations of the consistency relation 
\begin{equation}
F_{\rm AP}(z_\text{mean}) = F_{\rm th}(z_\text{mean}) \label{cons}
\end{equation}
indicates an inaccurate method for the estimation of the BAO scale. 

Although the fitting procedure, which requires a calibration, partially breaks the model independent assumption, the AP test offers the advantage of self-calibrating the fit 
by comparing parallel and transverse correlation functions.

We have performed several tests of our methodology: we tested the accuracy of
the AP test for different degrees  of the polynomial and different template
functions. The details of these tests are reported in Appendix~\ref{Ap:B}.
Our results show that the sufficient accuracy in the parameterization~(\ref{paramet}) is reached with $N=10$.

\section{Results}
\label{sec5}
In the previous section we summarized the method to recover 
the BAO peak position from the computed correlation function
in both radial and transverse directions.
In the next paragraphs we use this method to study 
possible sources that may affect the accuracy in the measurement of
the AP parameter $F_{\rm AP}$.

\subsection{RSD and lensing effects on the correlation function}
\label{sec5a}
In this section we aim to understand to which extent redshift space distortions and lensing corrections in the observed galaxy overdensity affect the measurement of the AP parameter. 
We compare the results of the AP test considering first  density perturbations only, and then  adding subsequently redshift space distortions and lensing corrections. These effects change the observed correlation function in radial and transverse directions, but clearly they do not change the intrinsic BAO scale.
For simplicity, in this section we assume the galaxies in our sample to be unbiased
tracers of the matter distribution, i.e.~$b = 1$. 

\begin{figure}[h!]
    \begin{center}  \vspace*{-1.5cm}
    \begin{subfigure}[b]{0.496\textwidth}
        \centering
        \caption{$\qquad z = 0.7$}
        \includegraphics[width=\textwidth]{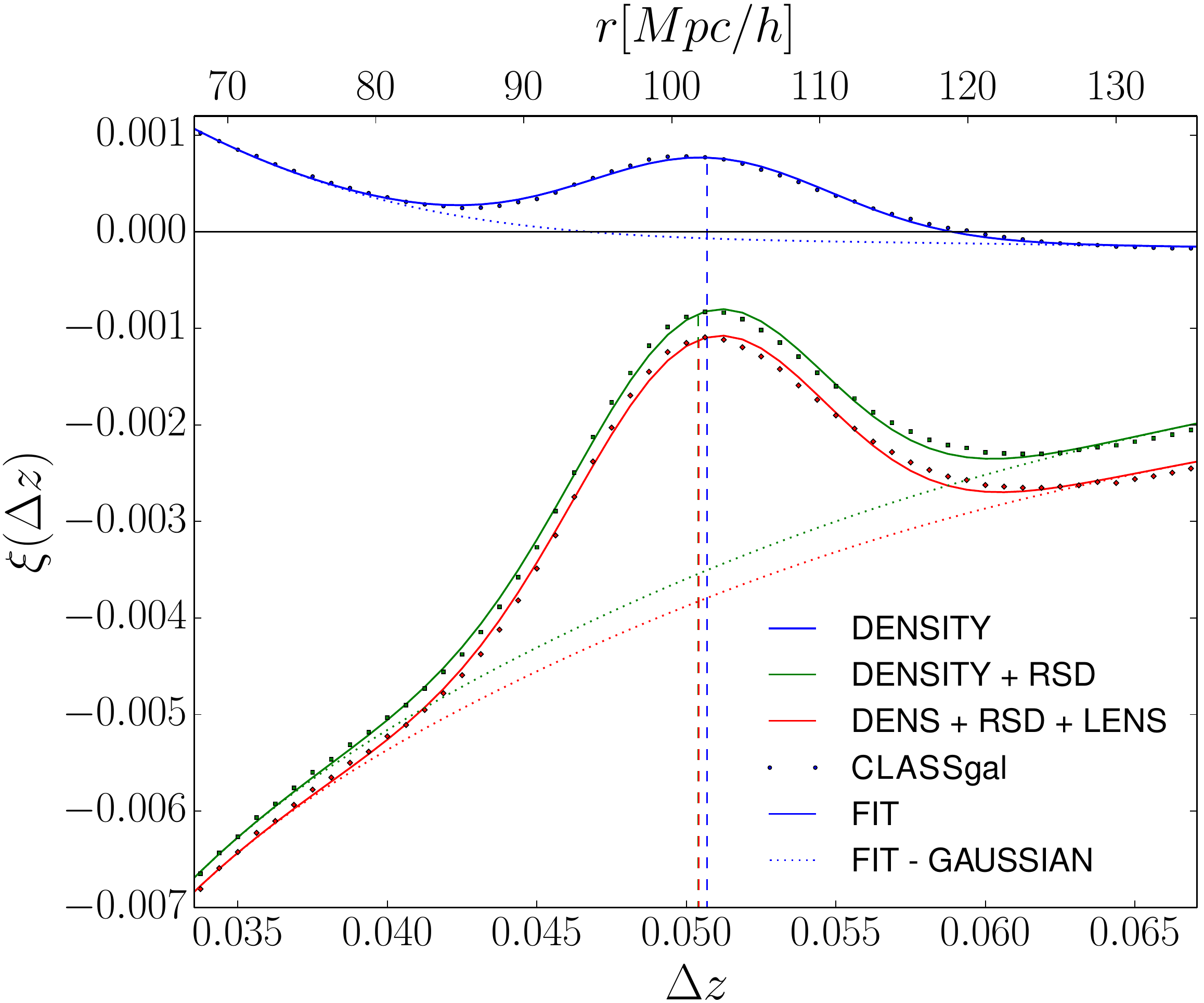}
        \label{}
    \end{subfigure}
    \hfill
    \begin{subfigure}[b]{0.496\textwidth}
     \caption{$\qquad z = 1.0$}
        \includegraphics[width=\textwidth]{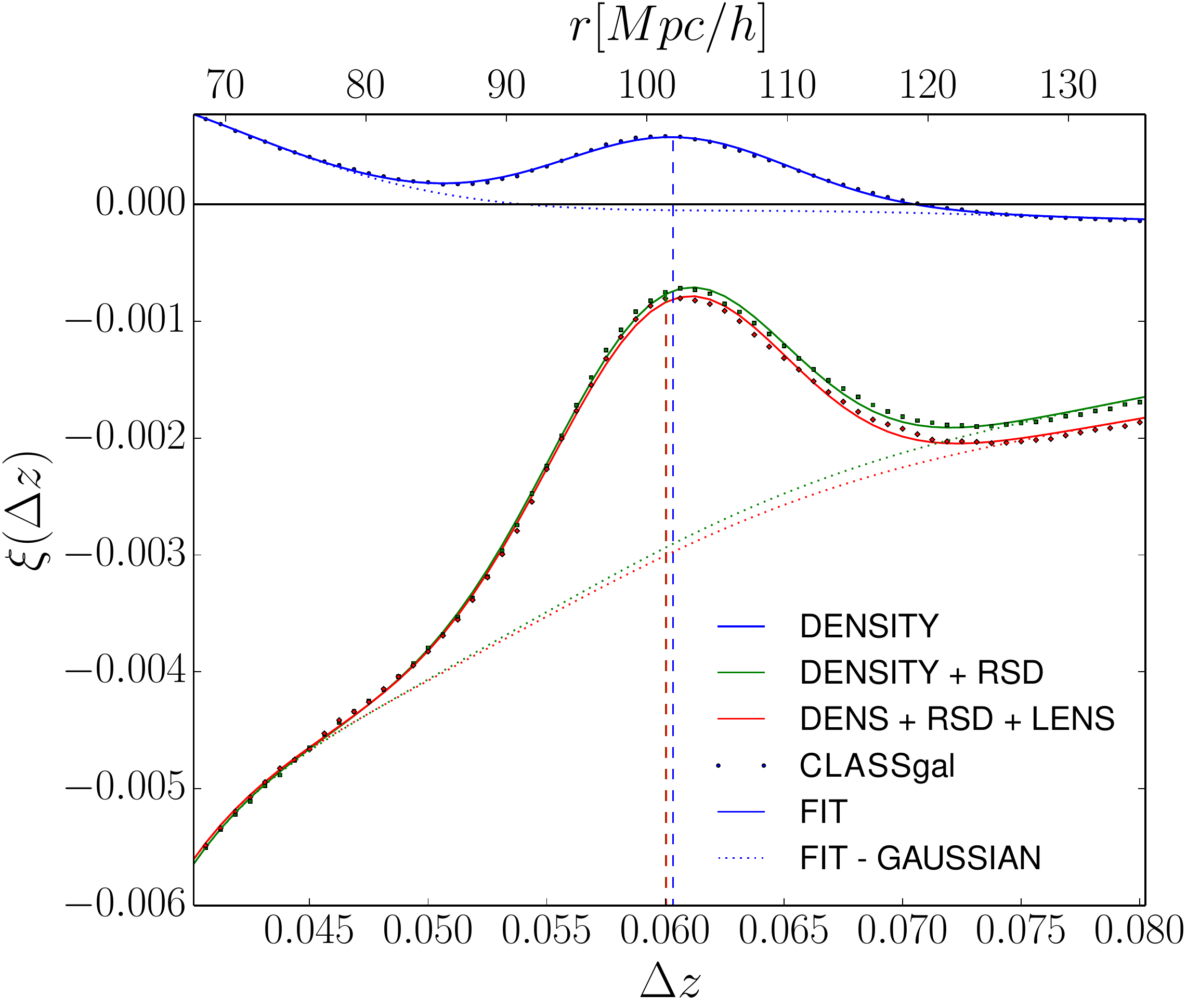}
        \label{}
    \end{subfigure}
    \\
    \begin{subfigure}[b]{0.496\textwidth}
        \caption{$\qquad z = 1.25$}
        \includegraphics[width=\textwidth]{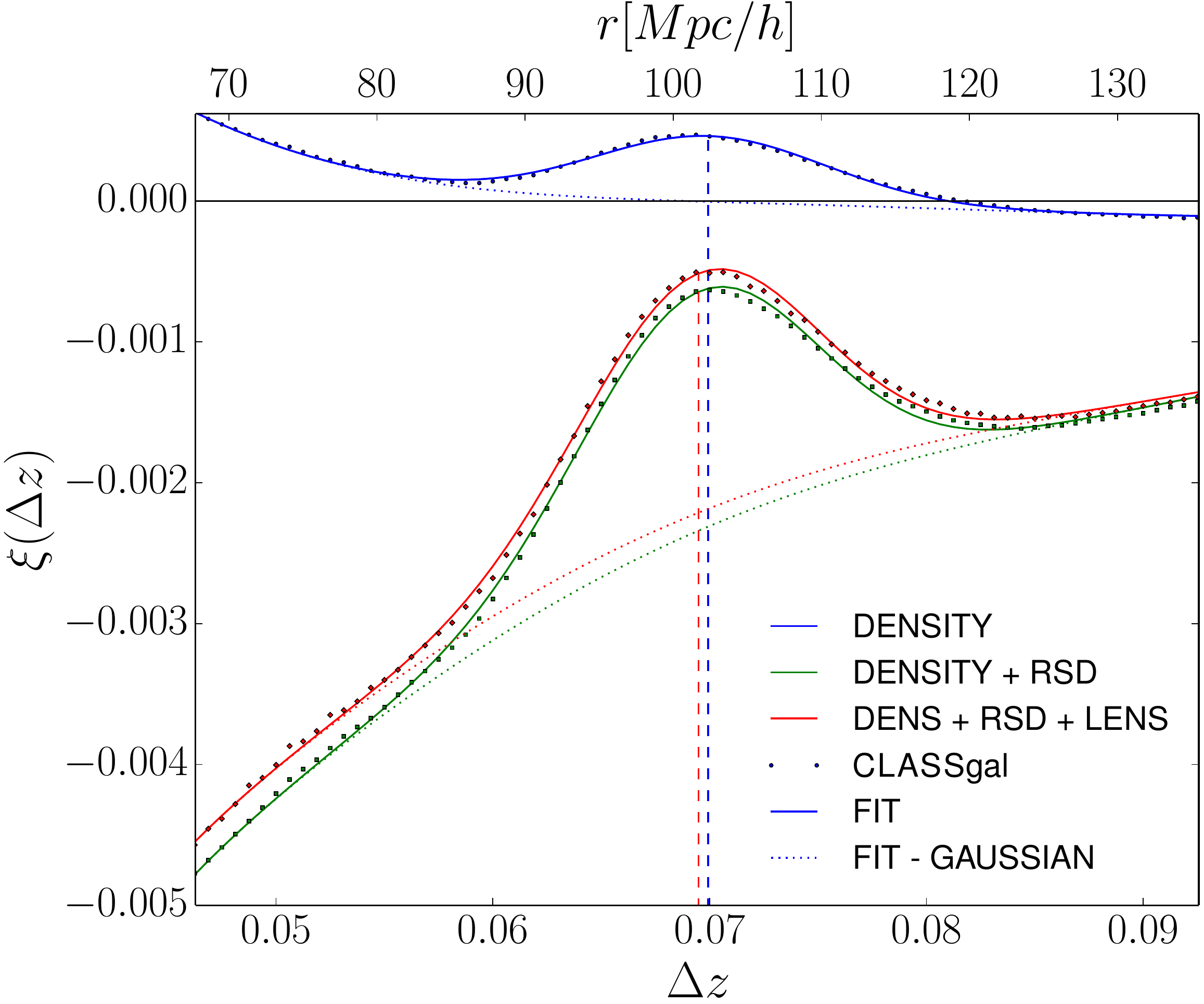}
        \label{fig:five over x}
    \end{subfigure}
    \hfill
    \begin{subfigure}[b]{0.496\textwidth}
        \caption{$\qquad z = 1.5$}
        \includegraphics[width=\textwidth]{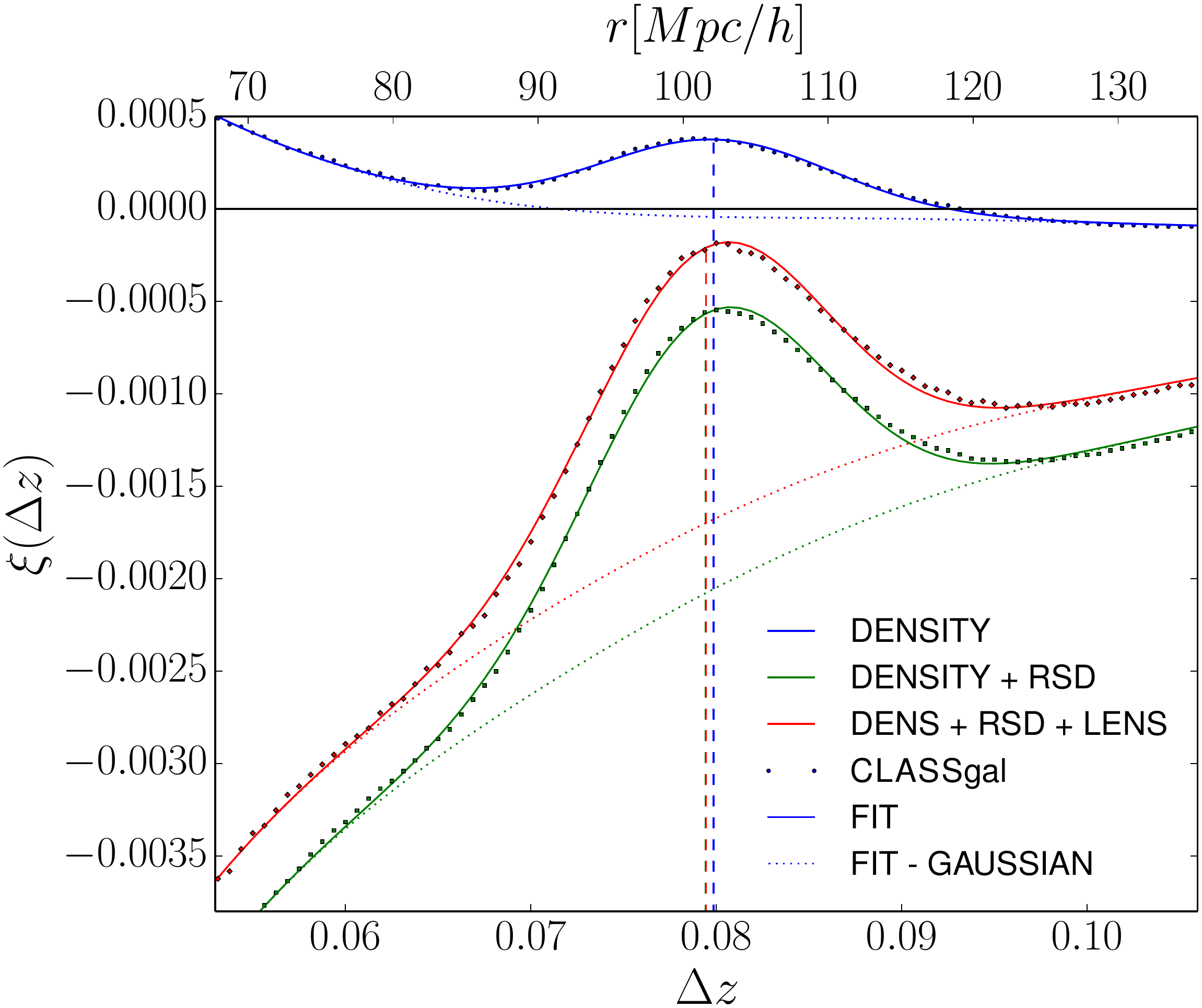}
        \label{}
    \end{subfigure}
    \\
    \begin{subfigure}[b]{0.496\textwidth}
    \begin{center}
       \caption{$\qquad z = 1.75$}
        \includegraphics[width=\textwidth]{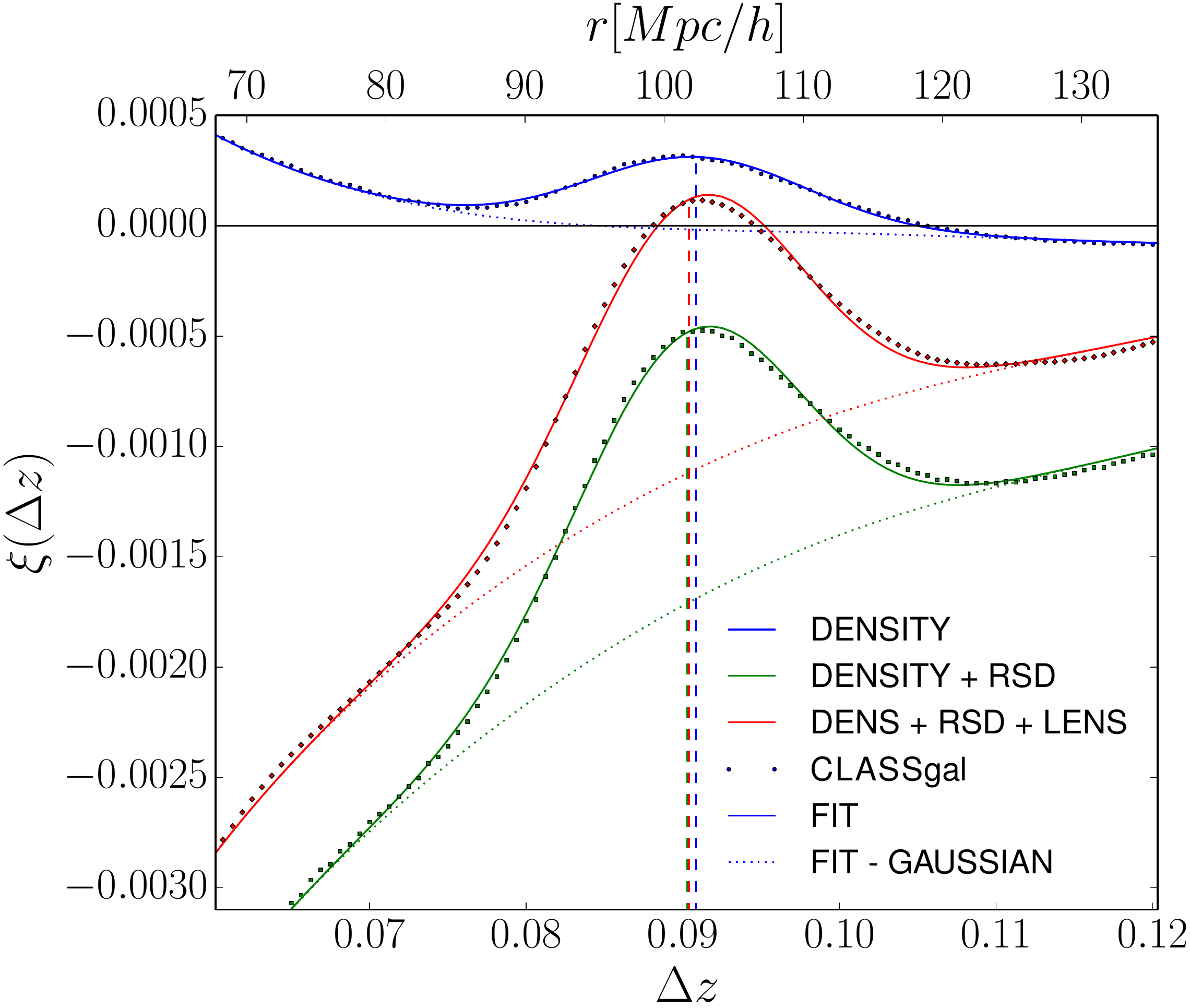}
        \label{}
     \end{center}
    \end{subfigure}
    \label{}
    \hfill
    \begin{subfigure}[b]{0.496\textwidth}
    \begin{center}
       \caption{$\qquad z = 2.0$}
        \includegraphics[width=\textwidth]{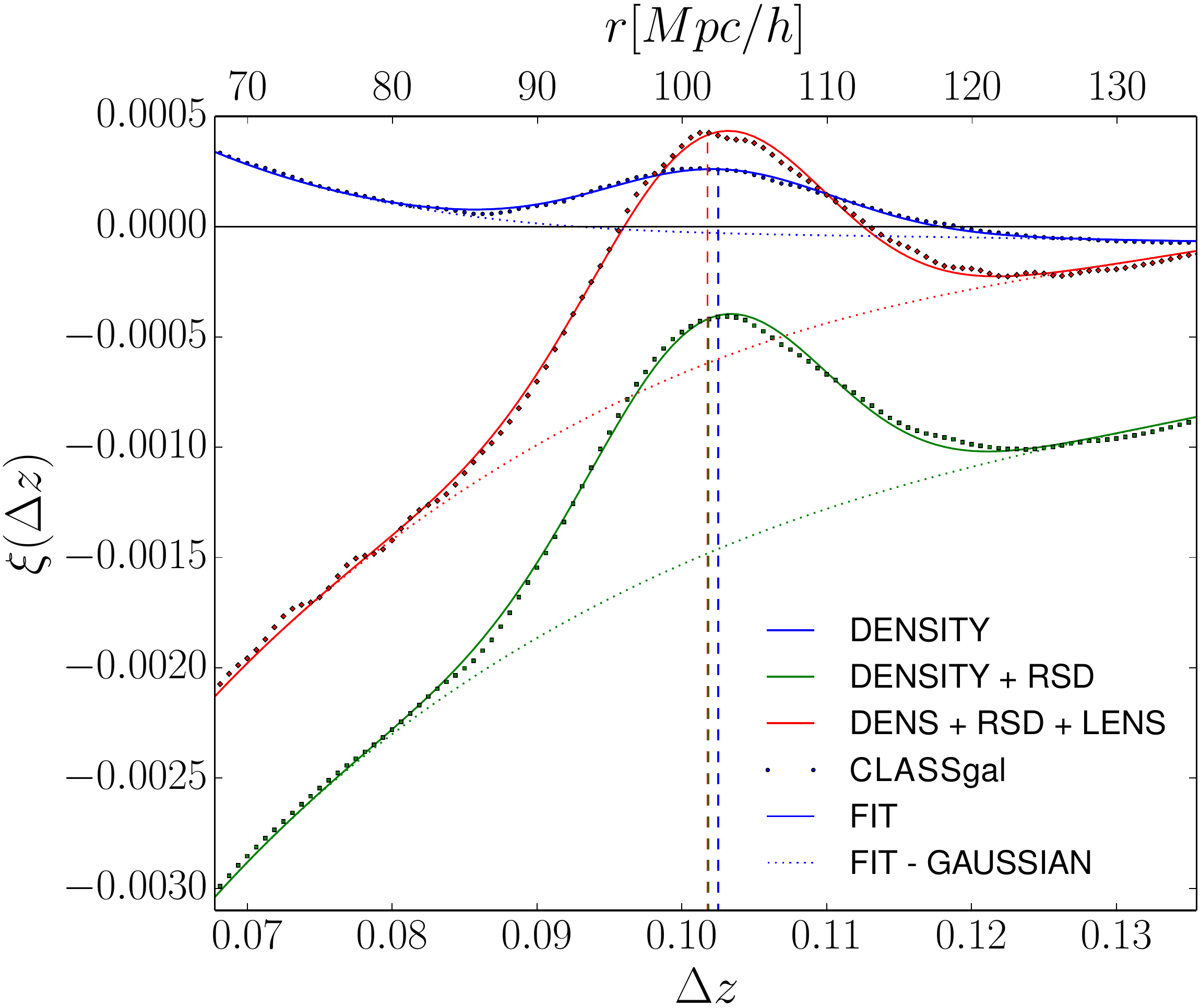}
        \label{}
     \end{center}
    \end{subfigure}
    \label{}
    \end{center}  \vspace*{-1.1cm}
\caption{The radial correlation function and the BAO peak position computed including different contributions to the observed over density.
The blue line refers to density correlations only,
the green line includes also the redshift space distortions, and the
red line takes into account  the previous  plus the lensing terms. The continuous lines refer to the best-fit model, while the dots are the {\sc class}gal output.
The dotted lines are obtained by subtracting the Gaussian term from the best-fit model.
The vertical dashed lines identify the estimated peak positions.
}
\label{Rad_fit}
\end{figure}

\begin{figure}[h!]
    \begin{center}
    \begin{subfigure}[b]{0.49\textwidth}
        \centering
        \caption{$\qquad z = 0.7$}
        \includegraphics[width=\textwidth]{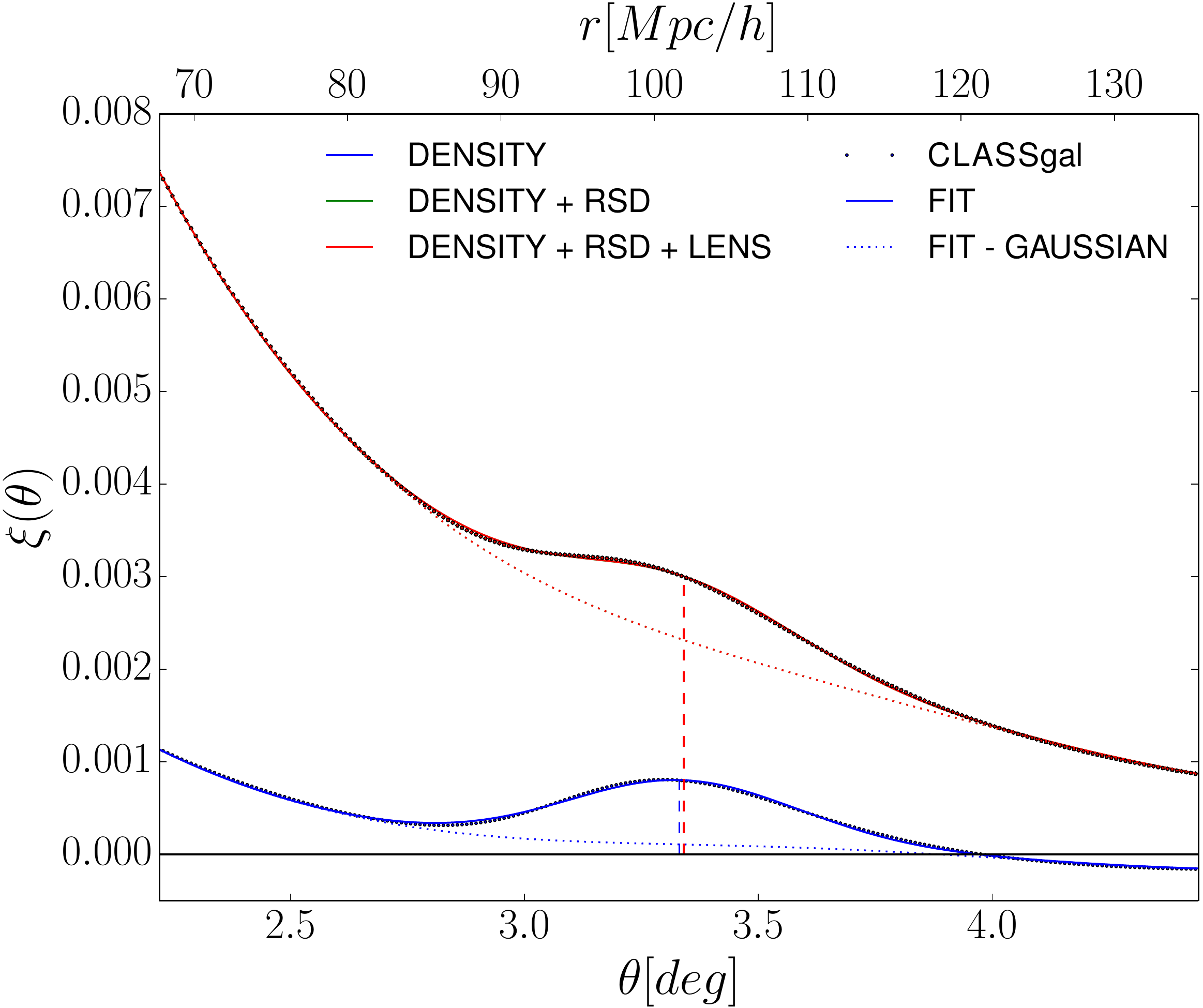}
        \label{}
    \end{subfigure}
    \hfill
    \begin{subfigure}[b]{0.496\textwidth}
               \caption{$\qquad z = 1.0$}
        \includegraphics[width=\textwidth]{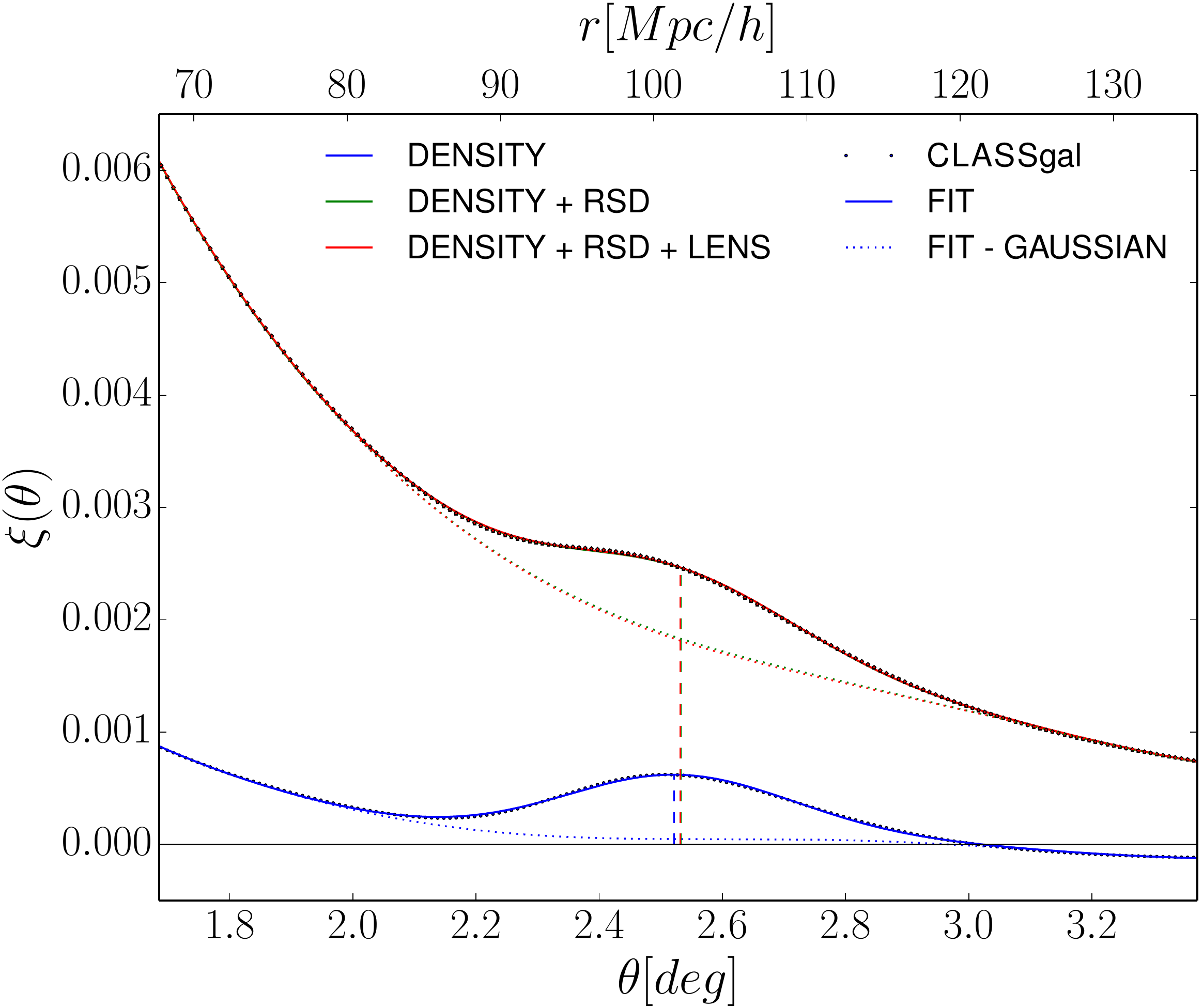}
        \label{}
    \end{subfigure}
    \\
    \begin{subfigure}[b]{0.496\textwidth}
        \caption{$\qquad z = 1.25$}
        \includegraphics[width=\textwidth]{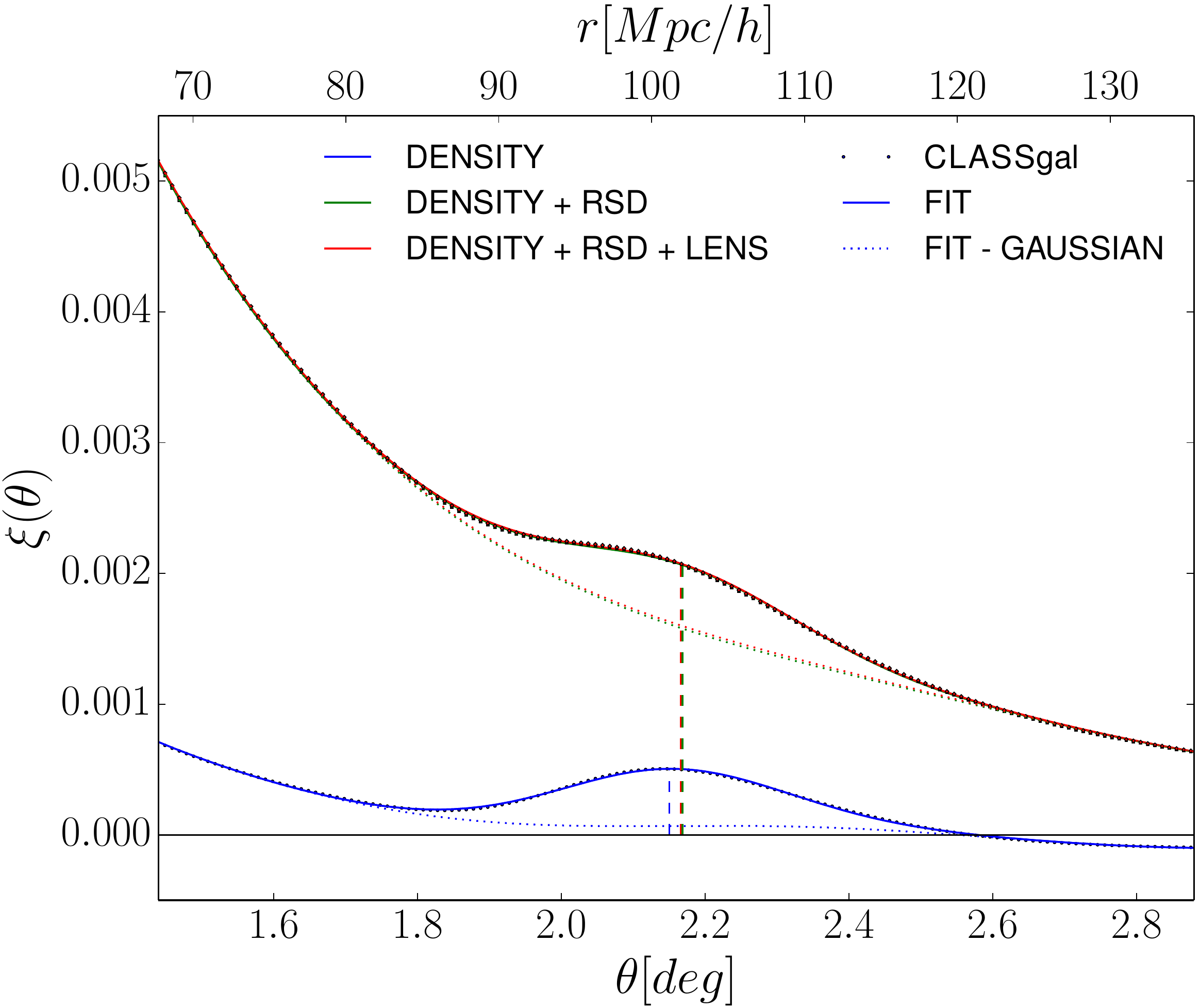}
        \label{fig:five over x}
    \end{subfigure}
    \hfill
    \begin{subfigure}[b]{0.496\textwidth}
          \caption{$\qquad z = 1.5$}
        \includegraphics[width=\textwidth]{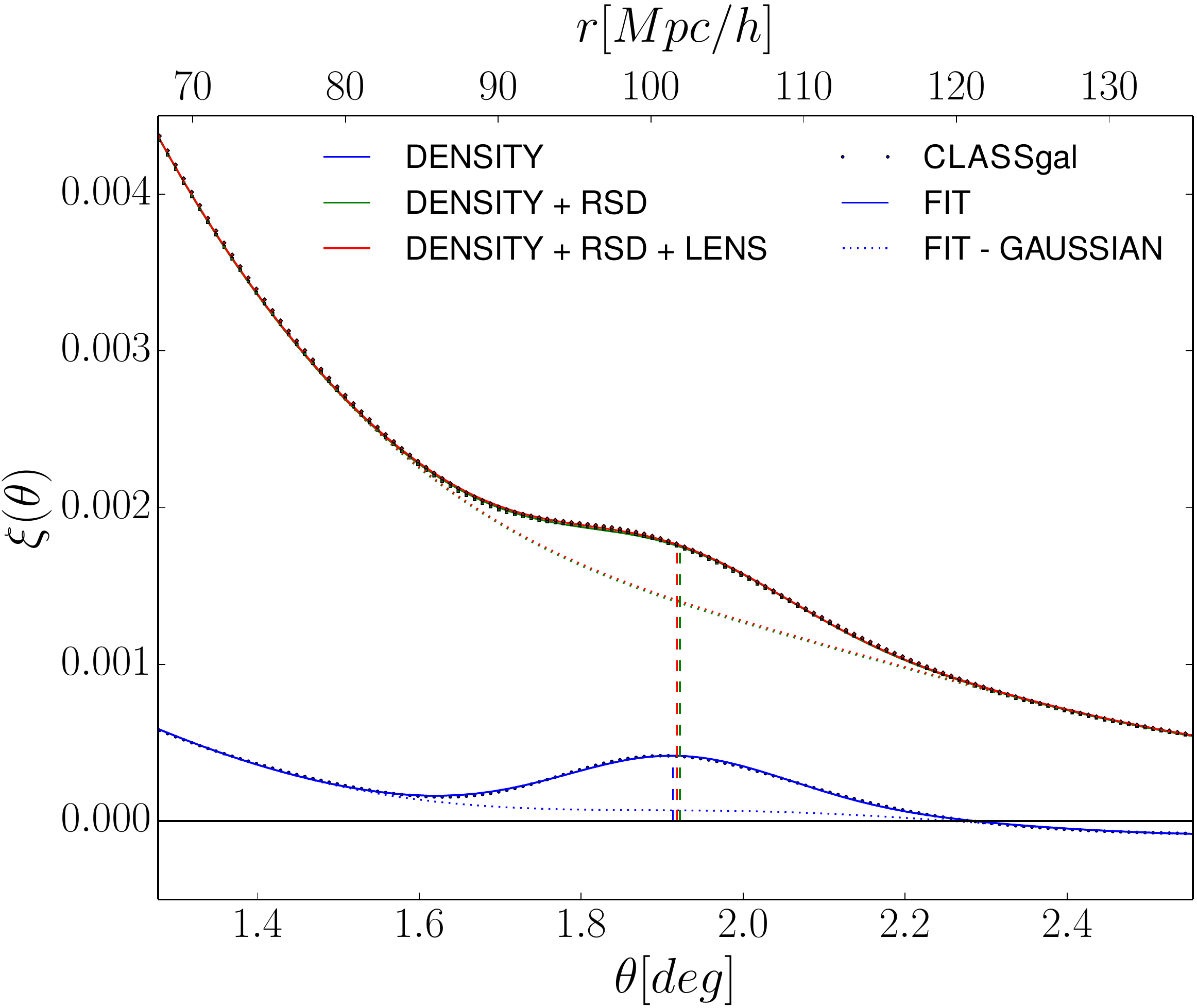}
        \label{}
    \end{subfigure}
    \\
    \begin{subfigure}[b]{0.496\textwidth}
    \begin{center}
         \caption{$\qquad z = 1.75$}
        \includegraphics[width=\textwidth]{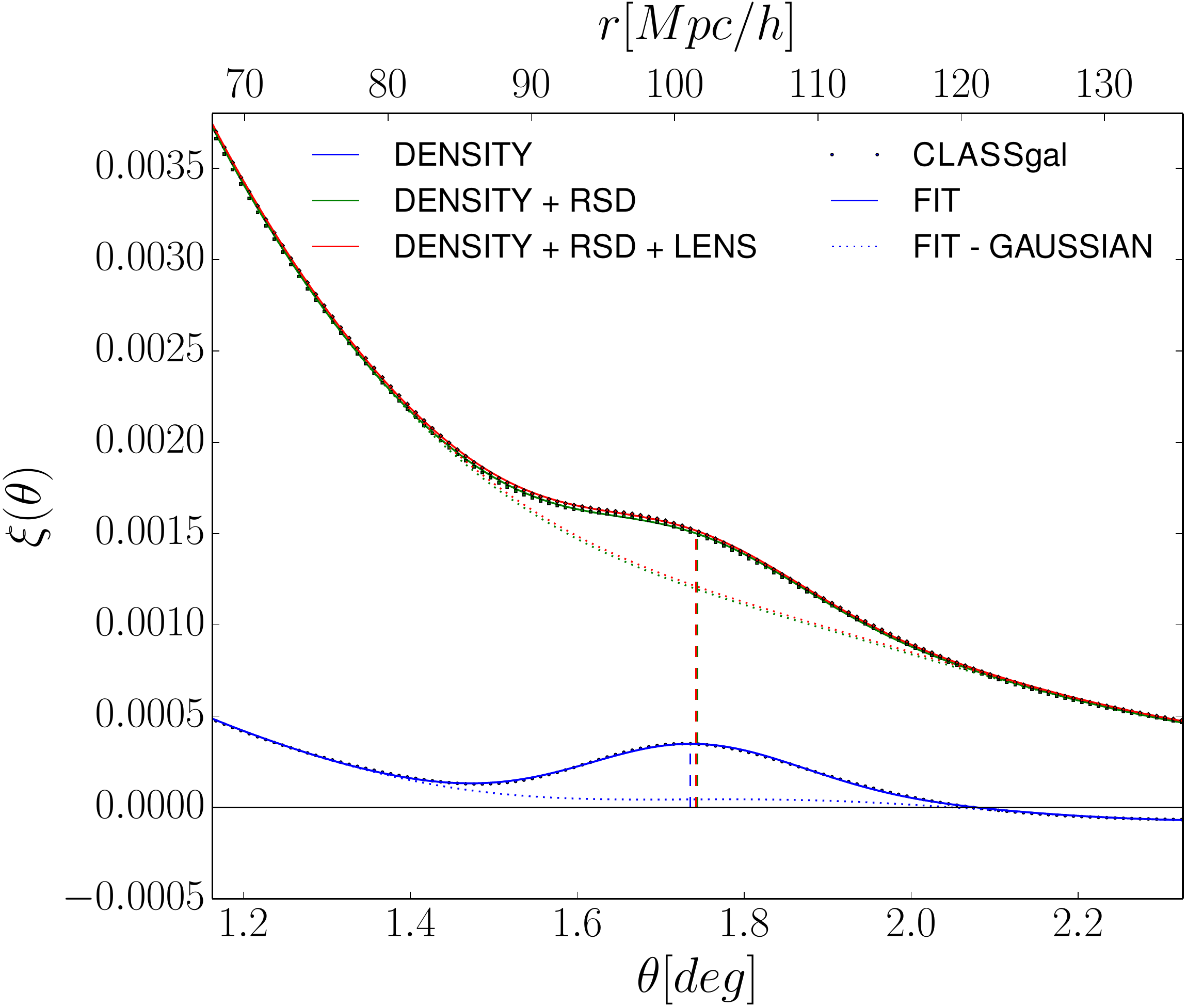}
        \label{}
     \end{center}
    \end{subfigure}
    \label{}
    \hfill
    \begin{subfigure}[b]{0.496\textwidth}
    \begin{center}
        \caption{$\qquad z = 2.0$}
        \includegraphics[width=\textwidth]{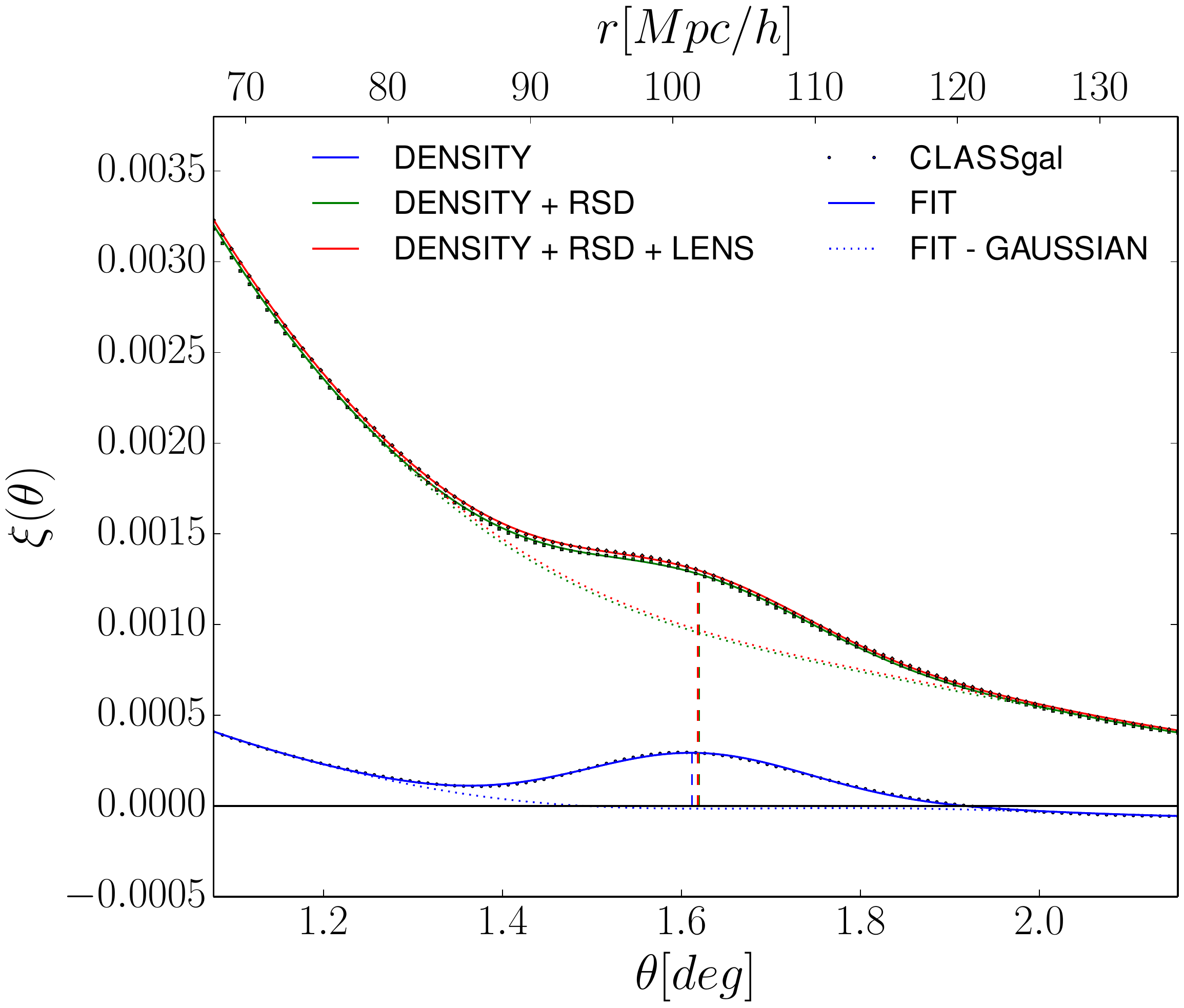}
        \label{}
     \end{center}
    \end{subfigure}
    \end{center}
    \vspace*{-1.1cm}
\caption{Transverse correlation function and BAO peak position computed including different contributions to the observed over density. 
We used the same colours and line styles as in Figure~\ref{Rad_fit}. \\
}
 \label{Transv_fit}
\end{figure}

Figure~\ref{Rad_fit} and Figure~\ref{Transv_fit} show the radial and the transverse correlations
and the corresponding fits for the three cases of interest, at
different redshifts, respectively.
As we have already seen in the previous sections, redshift space distortions significantly modify both the shape
and the amplitude of the correlation functions in the radial and transverse directions. We remark that the ratio between the density and the redshift space distortions terms is only weakly sensitive to the mean redshift. Indeed, in linear theory only the growth factor has a weak redshift dependence.
The estimated values of the position of the BAO peak are slightly
affected in both directions: when we do not include RSD corrections
the radial BAO peak is shifted toward larger scales, while
the angular peak position $\theta_{\rm BAO}$ is shifted
to smaller scales.
Although both effects are small ($\sim$ 0.5\%), they sum up
when applied to the AP test.
This result should not be interpreted as a physical shift of the acoustic scale,
it is a numerical effect related to the model assumed for the numerical fit,
which is calibrated to be valid in redshift space and not in real space.

\FloatBarrier

\begin{figure}[t!]
\begin{center}
\includegraphics[width=0.8\textwidth]{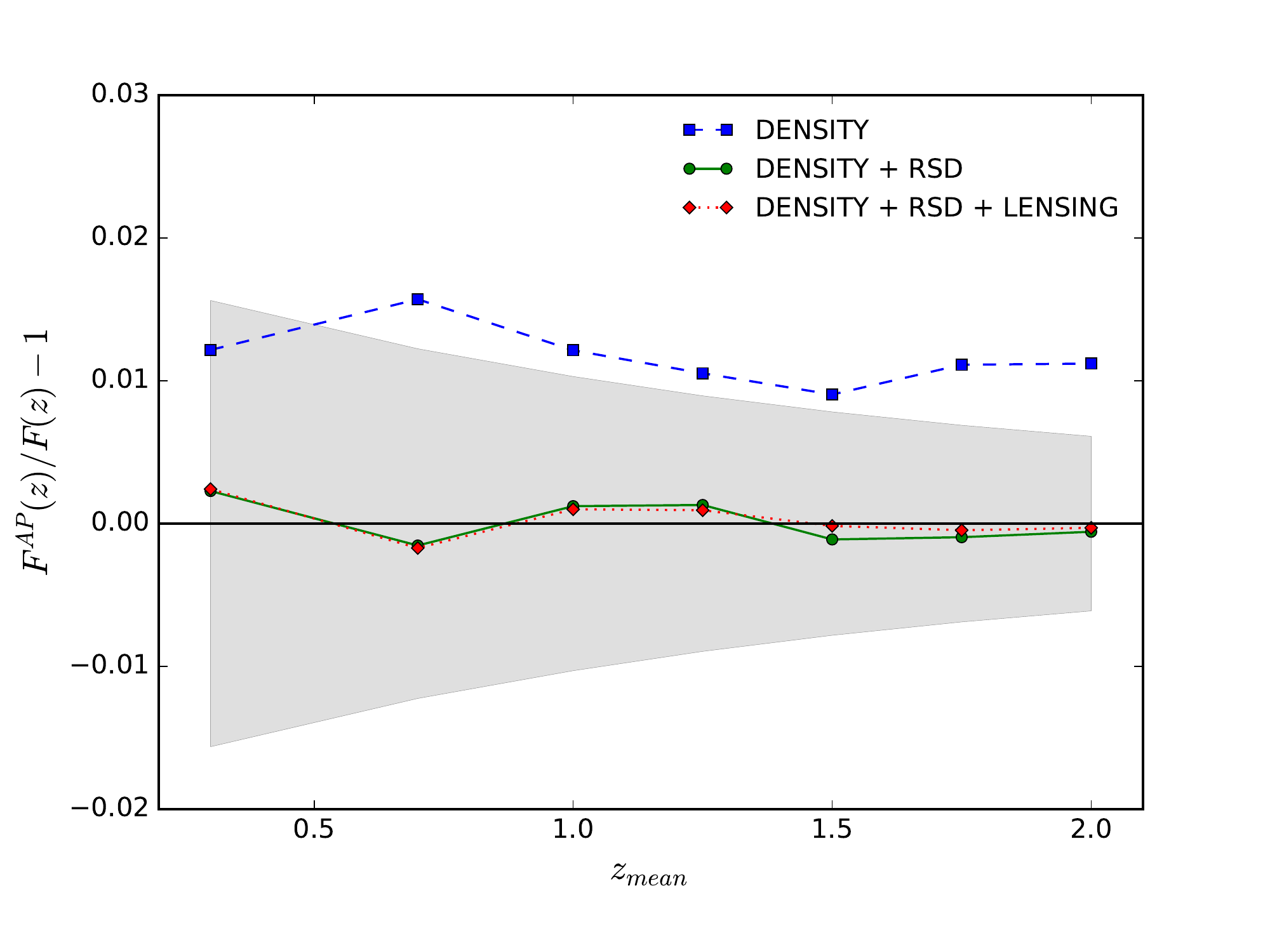}
\label{one}
\caption{AP consistency test for the three cases analysed in this section. The shaded region represents the expected errors around zero.
The errors on $F_{\rm AP}$ are estimated from the propagation of uncertainties, see eq. \eqref{prop_err} in Appendix \ref{Ap:B}.
Uncertainties on $\Delta z_{\rm BAO}$ and $\theta_{\rm BAO}$ are assumed
to be due to the resolution of the survey: we set $\delta z = 6.25 \cdot 10^{-4}$; this is  the resolution used to compute radial correlations,
while for the angular resolution we assume the resolution of 
a Euclid-like survey, i.e. $\delta \theta_{\rm BAO} = 0.1 \,\mbox{arcsec}$. The error on $F_{\rm AP}$ results to be dominated by the
uncertainty on $\Delta z_{BAO}$.}
\label{ap_gr}
\end{center}
\end{figure}

The gravitational lensing correction does not affect the transverse correlation function, while in the radial direction it changes the amplitude, but not the shape of the correlation function. Therefore, it does
not appreciably shift the peak position. In Figure~\ref{Rad_fit} we show that the lensing term reduces the amplitude of the radial correlation function at low redshifts, while it increases the amplitude at high redshifts. This is due to the fact that the lensing contribution is dominated by the negative correlation between density and lensing at low redshifts, while is dominated by the auto-correlation at high redshifts.

In Figure~\ref{ap_gr} we show the result of the consistency test 
for the three cases of interest here.
We find that gravitational lensing does not affect
the AP test, while redshift space distortions enhance the result by about 1\% if compared to an analysis performed in real space.

\subsection{Galaxy bias}
\label{sec5b}
In the previous section we assumed observed galaxies to be unbiased tracers of the underlying dark matter distribution. 
In this section we relax this assumption in order to investigate the implication
of galaxy bias for the BAO measurements and for the AP test. 
The effect of galaxy bias on BAO measurement was investigated 
in~\cite{bias}, where it was found that for the most biased tracers
($b > 3$) a non-linear shift on the acoustic scale occurred at the percent level.
However, it has been  demonstrated that applying
BAO reconstruction~\cite{reconstruction} compensates
for this effect. 

In this work we aim to study how a simple
local and linear bias model influences the AP test.
In fact, the bias affects the computation of the correlation function,
but does not enter in the theoretical value of the AP parameter. 
Hence, any deviation from the theoretical value, is due to our ability of determining the BAO scale in terms of the truly observable quantities $\xi_\parallel \left( \Delta z, z_\text{mean} \right)$ and $\xi_\perp \left( \theta, z_\text{mean} \right)$.
Since in the previous section we have found that lensing does not affect the 
measurement of the AP parameter, we neglect the
lensing contribution here.
We follow an analogous procedure as in the previous section, but here
 we vary the bias parameter $b$, that we assume to be redshift and scale independent (this assumption can also be easily generalized).

\begin{figure}[t]
\begin{center}
\includegraphics[width=0.8\textwidth]{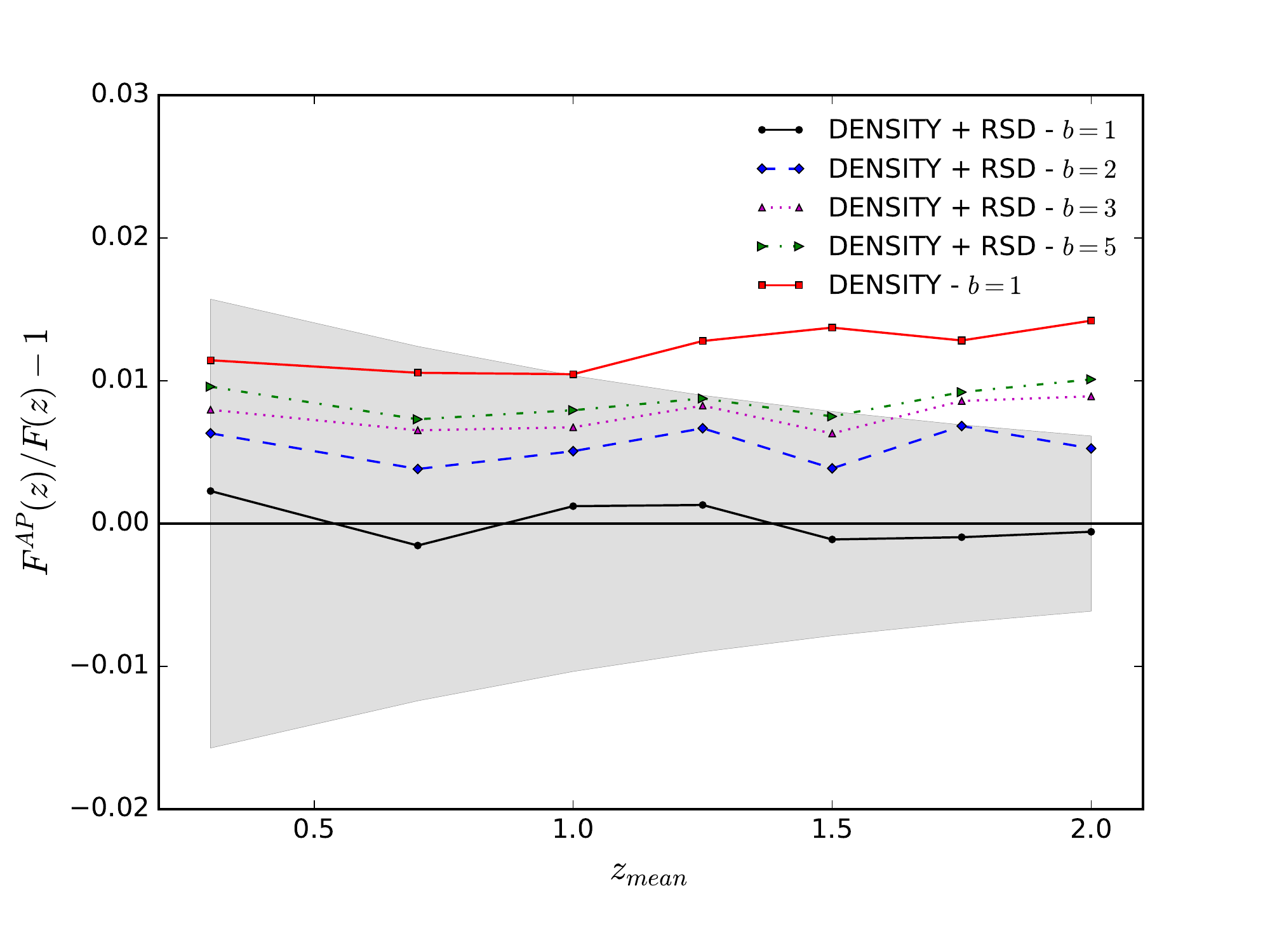}
\label{one}
\caption{AP consistency test for different values of the bias parameter $b$. The red line is computed including the only contribution from the local density field. The shaded region highlights the typical error-bars around zero.
The most biased tracers cause an offset up to $1\%$. This systematic uncertainty should be taken into account in the total error budget.
}
\label{ap_bias}
\end{center}
\end{figure}

In Figure~\ref{ap_bias} we show the result of the AP consistency test for different values of the bias parameter.
We see that the bias causes an offset which grows with increasing bias. This offset is due to the fact that the BAO position is recovered by using the phenomenological parameterization~(\ref{paramet}), which has been calibrated for unbiased sources. From Figs.~{\ref{Rad_fit} and~\ref{Transv_fit}} we note that the shape changes considerably when including redshift space distortions in the correlation function. Therefore, correlation functions for different galaxy bias parameters ($b>1$) range between the blue and the green lines in Figs.~{\ref{Rad_fit} and~\ref{Transv_fit}} and they affect the precision of the parameterization. This indicates, that even when working only with directly observable quantities, we need to assume some cosmological prior to be able to determine accurately the BAO position. Nevertheless, to perform an AP test we do not need to know the physical scale of BAO, and so we can use radial and transverse correlation functions to self-calibrate the parameterizations.

\begin{figure}[t!]
    \centering
    \begin{subfigure}[b]{0.6\textwidth}
        \centering
        \includegraphics[width=\textwidth]{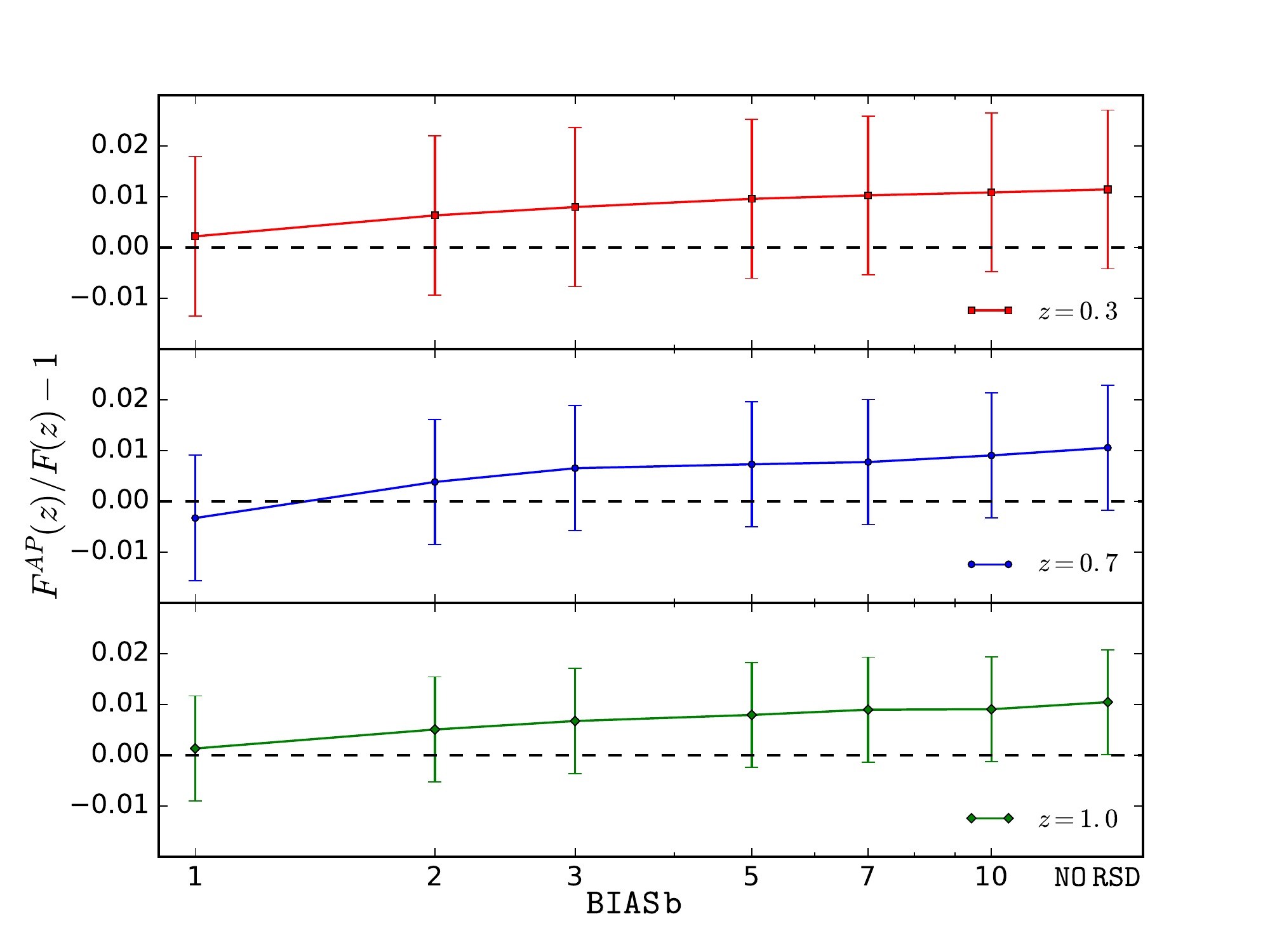}
    \end{subfigure}
     %\hfill
    \begin{subfigure}[b]{0.6\textwidth}
        \includegraphics[width=\textwidth]{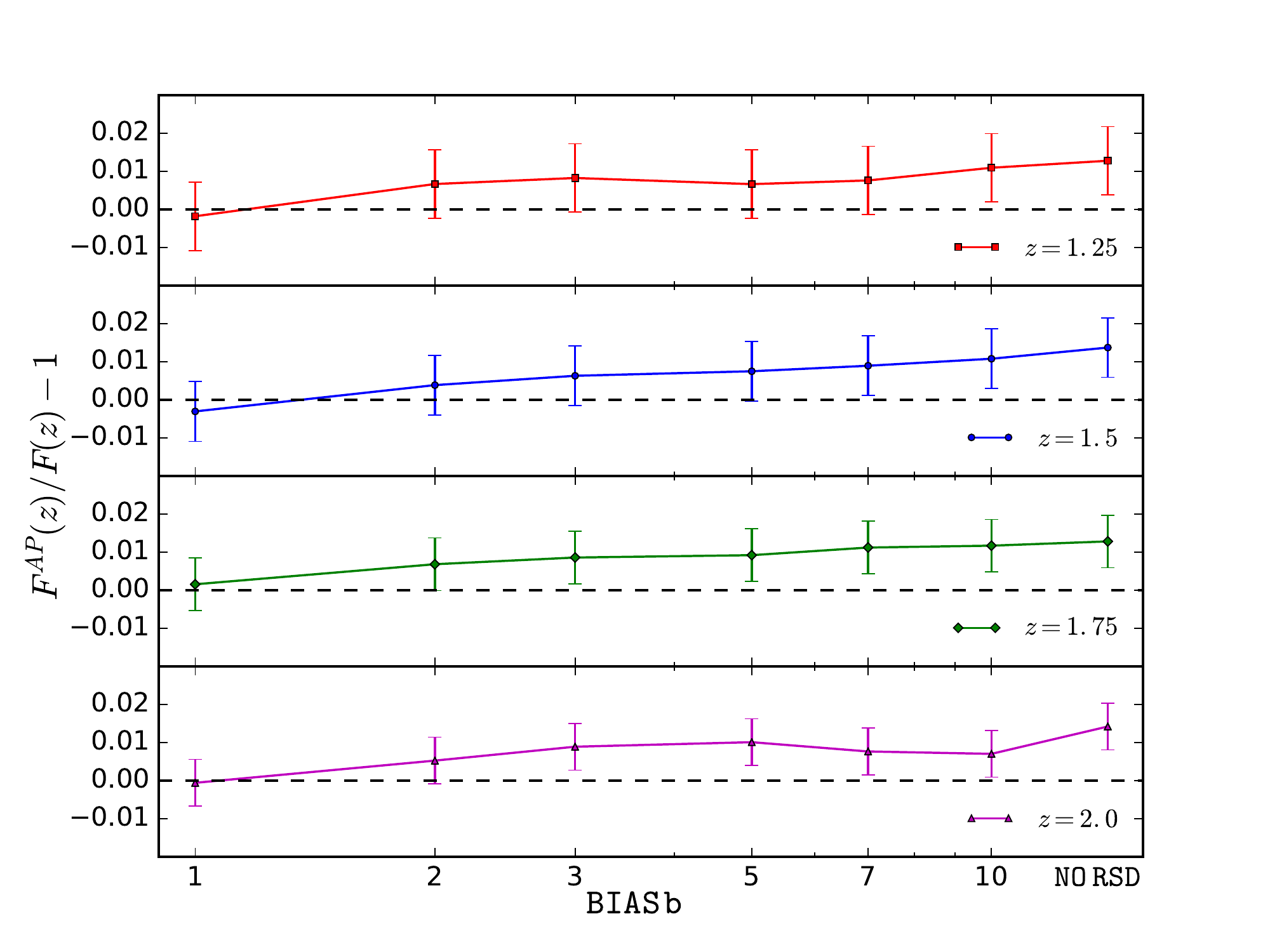}
    \end{subfigure}
\caption{AP consistency test, at fixed redshift, for different values
of the bias parameter. 
\\
}
\label{ap_bias_2}
\end{figure}

In Figure~\ref{ap_bias_2} we show how the offset varies as a  function
of the bias parameter, at fixed redshift.
When increasing the bias, this offset approaches the offset 
found for an unbiased tracer, when only the local density term is
taken into account. Indeed, for large galaxy bias parameters the redshift space distortions contribution tends to be negligible. As expected there is only a marginal redshift dependence, since we have considered a redshift independent bias and there is only a weak redshift dependence in the relation between the density and the velocity transfer functions, given by the growth factor. 
In a given survey, it is therefore important to reconstruct the bias as good as possible.
In this way the fitting procedure can be calibrated for the estimated bias and the impact of the systematic error investigated in this section can be mitigated.

\subsection{The radial window function}

\label{window}

In the previous section we assumed that the redshifts of the sources
are exactly known. In realistic applications, though, a redshift bin
has finite thickness that can be modeled by a window function
(usually a Gaussian or a top-hat) centered at some mean redshift.

To be able to locate the radial BAO peak position a spectroscopic precision (typically $\sigma_z = 0.001(1+z_\text{mean})$ or smaller) for the redshift determination is required, whereas photometric redshift resolution is not sufficient. 

For this reason we will focus here on spectroscopic surveys.

\begin{figure}[t]
\begin{center}
\includegraphics[width=0.8\textwidth]{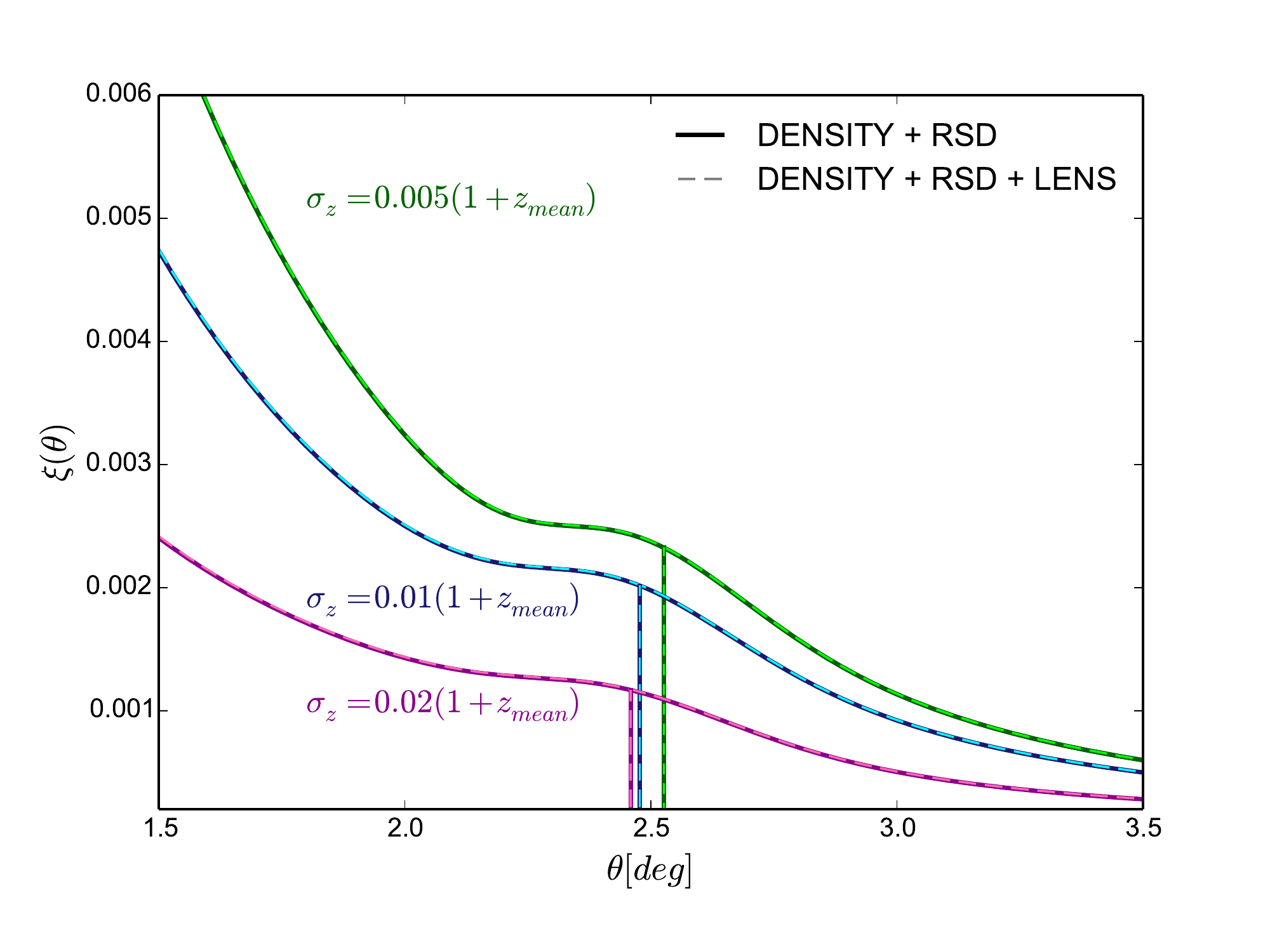}
\caption{Transverse correlation function, at $z=1$, for different width of the window function. For the thick continuous lines 
only density and redshift space distortions are included, while in the thin dashed lines also the lensing term is taken into account.}
\label{wind_lens}
\end{center}
\end{figure}

However, in the transverse correlation function, the presence of a radial window function does not affect the resolution, but it smears out the BAO feature. As shown in Fig.~\ref{wind_lens}, redshift bins of a typical width of photometric survey are sufficient to locate the peak position, but the position depends on the bin-width,  see also~\cite{montanari_durrer}. 

Even if the AP test can be performed only with data from a spectroscopic survey, we can employ a window function with width larger than the redshift uncertainty in order to maximize the signal-to-noise ratio
\begin{equation}
\Biggl(\frac{S}{N}\Biggr)_{\theta} = \frac{\xi_{\perp}(\theta)}{\sigma_{\xi_{\theta}}},
\end{equation}
where the noise $\sigma_{\xi_{\theta}}$ is given by
the root mean square of the diagonal elements of the covariance
matrix $\mathtt{COV}$
\begin{gather}
\mathtt{COV}_{\theta\theta'} =\frac{2}{f_{\rm sky}} \sum_{\ell=0}^{\ell_{max}} \frac{2\ell + 1}{(4\pi)^2}\Bigl[P_{\ell}(\cos{\theta})\Bigr] \Bigl[P_{\ell}(\cos{\theta'})\Bigr]  \Biggl(C_{\ell} + \frac{1}{n_i}\Biggr)^2, \label{covariance} \\
\sigma_{\xi_{\theta}} = (\mathtt{COV}_{\theta\theta})^{1/2}.  \label{err}
\end{gather}
In the expression \eqref{covariance}, 
$f_{\rm sky}$ is the fraction of the sky covered 
by the survey.
The shot noise is 
computed as the inverse of
the number of galaxies per steradian $n_i$ inside the $i$-th redshift bin.
  
In order to find the width of the window function which maximizes the signal-to-noise ratio, we compute 
this quantity for different values of $\sigma_z$
\begin{equation}
\sigma_z = \mathtt{w} \cdot (1+z_{\rm mean}) , \qquad \mathtt{w} = 0.005, 0.01, 0.02, 0.03,
\end{equation}
at different redshift in the range $z_{\rm mean} =  [0.7-2.0]$ and for the different specifications of the survey.
%---------------------------------------------------------------------------------------------------------------------------
We consider  three future planned galaxy redshift surveys: Euclid, the Square Kilometer Array (SKA) and
the Dark Energy Spectroscopic Instrument (DESI).
As for DESI, we consider the two classes of galaxies that will be
targeted by this experiment: bright Emission Line Galaxies (ELGs),
that will be observed up to $z=1.7$,
and Luminous Red Galaxies (LRGs), that will be observed up to $z=1.0$. 
%---------------------------------------------------------------------------------------------------------------------------

In Appendix~\ref{AP:C} we present the survey specifications we use
in the computation of the signal-to-noise ratio.
%%%%%%%%%%%%%%%%%%%
\begin{figure}[t!]
    \begin{subfigure}[b]{0.5\textwidth}
        \centering
        \includegraphics[width=\textwidth]{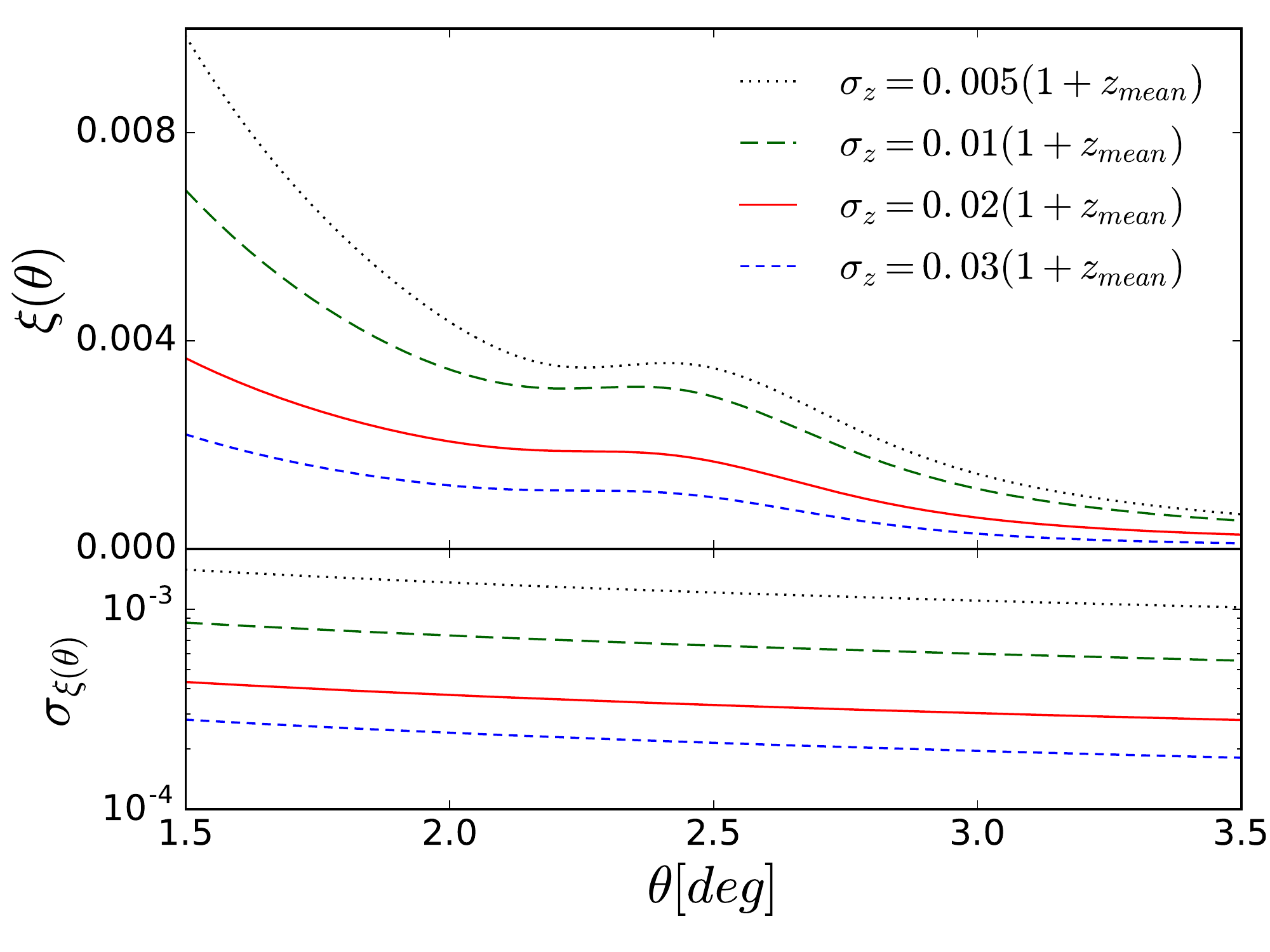}
         \caption{Transverse correlation function and noise.}
        \label{SN_A}
    \end{subfigure}
    \begin{subfigure}[b]{0.5\textwidth}
        \includegraphics[width=\textwidth]{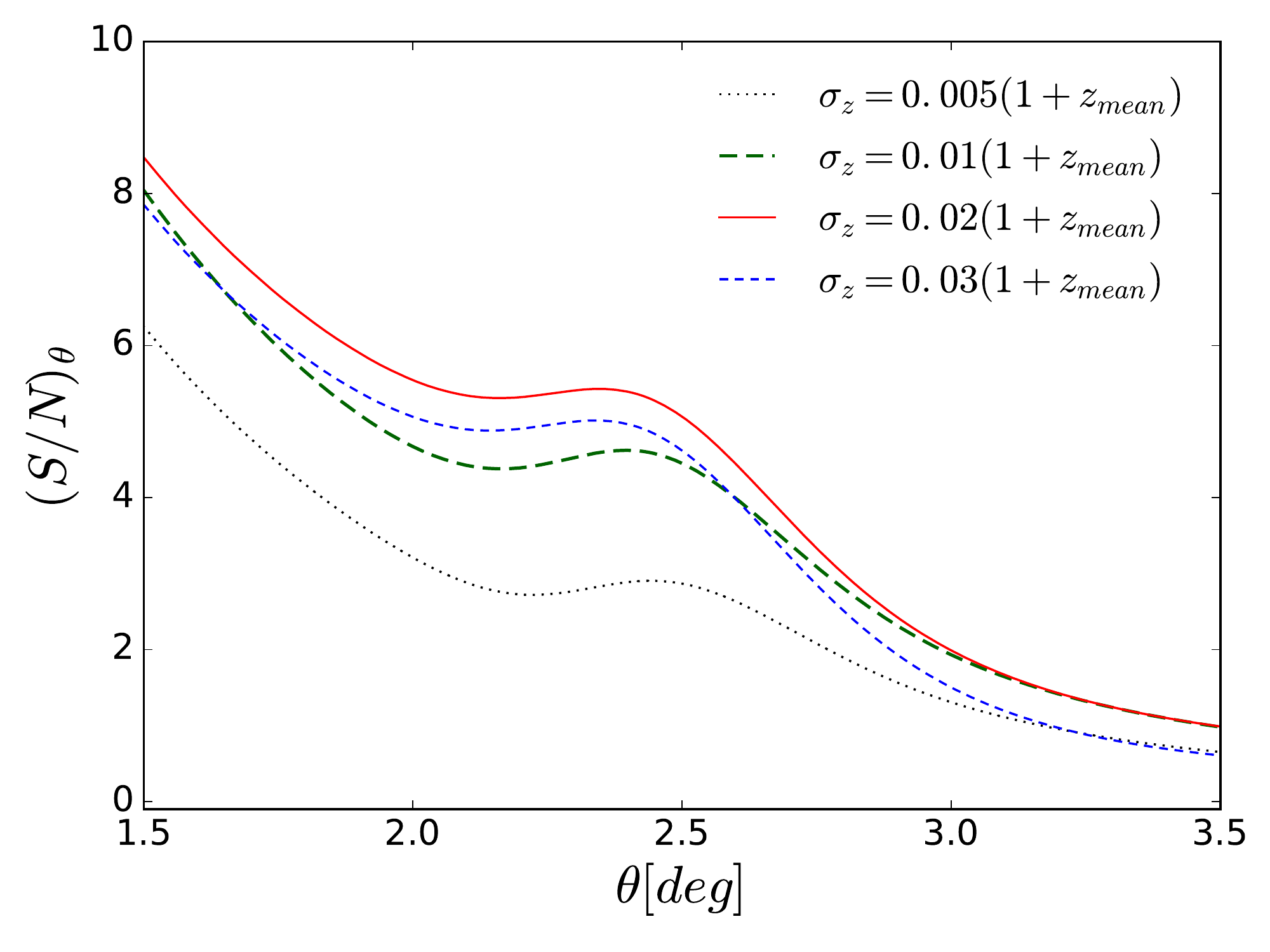}
     \caption{Signal-to-noise ratio.}
    \label{SN_B}
    \end{subfigure}
\caption{In Figure~\subref{SN_A} and Figure~\subref{SN_B} we show,
respectively, the transverse correlation function with its corresponding noise and the signal-to-noise ratio for different widths of the Gaussian
window function. Both figures refer to the case of an Euclid-like
survey at redshift $z=1$.} 
\label{SN}
\end{figure}
%%%%%%%%%%%%%%%%
In Figure~\ref{SN} we show, as an example, 
the comparison of the signal-to-noise ratio for a
Euclid-like galaxy survey at redshift $z = 1$.
From Figure~\ref{SN_A} we see that, for a large window function, the
signal is suppressed, but the correspondent noise also decreases.
We find that the signal-to-noise ratio, around the BAO peak, 
is optimized for $\sigma_z = 0.02(1+z_{\rm mean})$ (Figure~\ref{SN_B}).
For all the four sets of survey specifications analyzed here, we find the same result. Therefore, from now on, we
set our window function to $\sigma_z = 0.02(1+ z_{\rm mean})$.
%%%%%%%%%%%%%%%%%%%%%%%%%%%%%%%%%%%%%%%%%%%%%%%%
%%%  

Now we are interested in studying the effect of a 
window function on the transverse correlation function and thereby on the AP test. 

Since we consider window function with width much larger than the redshift resolution,
we assume a top-hat radial window in the transverse
correlation function and we refer to $\sigma_z$ as the half-width of the top-hat. 

In the radial correlation, a window with spectroscopic width
($\sigma_z = 0.001(1+z_\text{mean})$ or smaller) does not substantially 
affect the correlation function because it still includes only radial modes.
Therefore, we used a Delta Dirac window. 

In Figure~\ref{wind_lens} we compare the transverse correlation function
computed applying three window functions of different widths.
In this example, we assume bias $b=1$ and, 
use the analytic expression from~\cite{dn_dz} for the galaxy density distribution,
\begin{equation}
\frac{dN}{dz}\propto \Biggl(\frac{z}{0.5}\Biggr)^2 \exp{\Biggl[-\biggl(\frac{z}{0.5}\biggr)^{1.5}\Biggr]}.
\end{equation}
The window function suppresses the correlations and it shifts
 the peak position toward smaller angular scales, which is in agreement with the result presented in~\cite{montanari_durrer}.
Furthermore, we test the modifications induced by gravitational lensing
in the transverse direction when a window function is employed.
We find that gravitational lensing slightly modifies the amplitude 
of the correlation function, but this effect is negligible with respect
to the suppression induced by the window and leaves the peak position unaffected. 

In order to recover the physical BAO scale, 
the shift induced by the window function can be modeled as longitudinal component in the measured BAO scale.
More precisely, the BAO scale can be computed as
\begin{equation}
L = \sqrt{L^2_{\perp} + \Biggl(\frac{\delta z}{H(z)}\Biggl)^2},
\label{correct}
\end{equation}
where $L_{\perp}$ is estimated from the BAO peak in transverse
correlation function, while the corrective term $\delta z$
is a function of the width $\sigma_z$ of the window function.
We use the parameterization introduced in~\cite{montanari_durrer}, where the corrective term is simply
proportional to $\sigma_z$
\begin{equation}
\delta z = \sqrt{\gamma} \sigma_z.
\label{deltaz_eq}
\end{equation}  

Here we predict
the value of the corrective factor $\gamma$ from the AP test, 
for three future galaxy redshift surveys previously quoted: Euclid, SKA and DESI.

We introduce also a realistic model for the redshift dependence of
galaxy bias.

We perform the AP test, by introducing the radial window in the transverse correlations and
modeling a redshift dependent bias in both radial and transverse
correlation functions.

\begin{figure}[!ht]
\begin{center}
\includegraphics[width=0.8\textwidth]{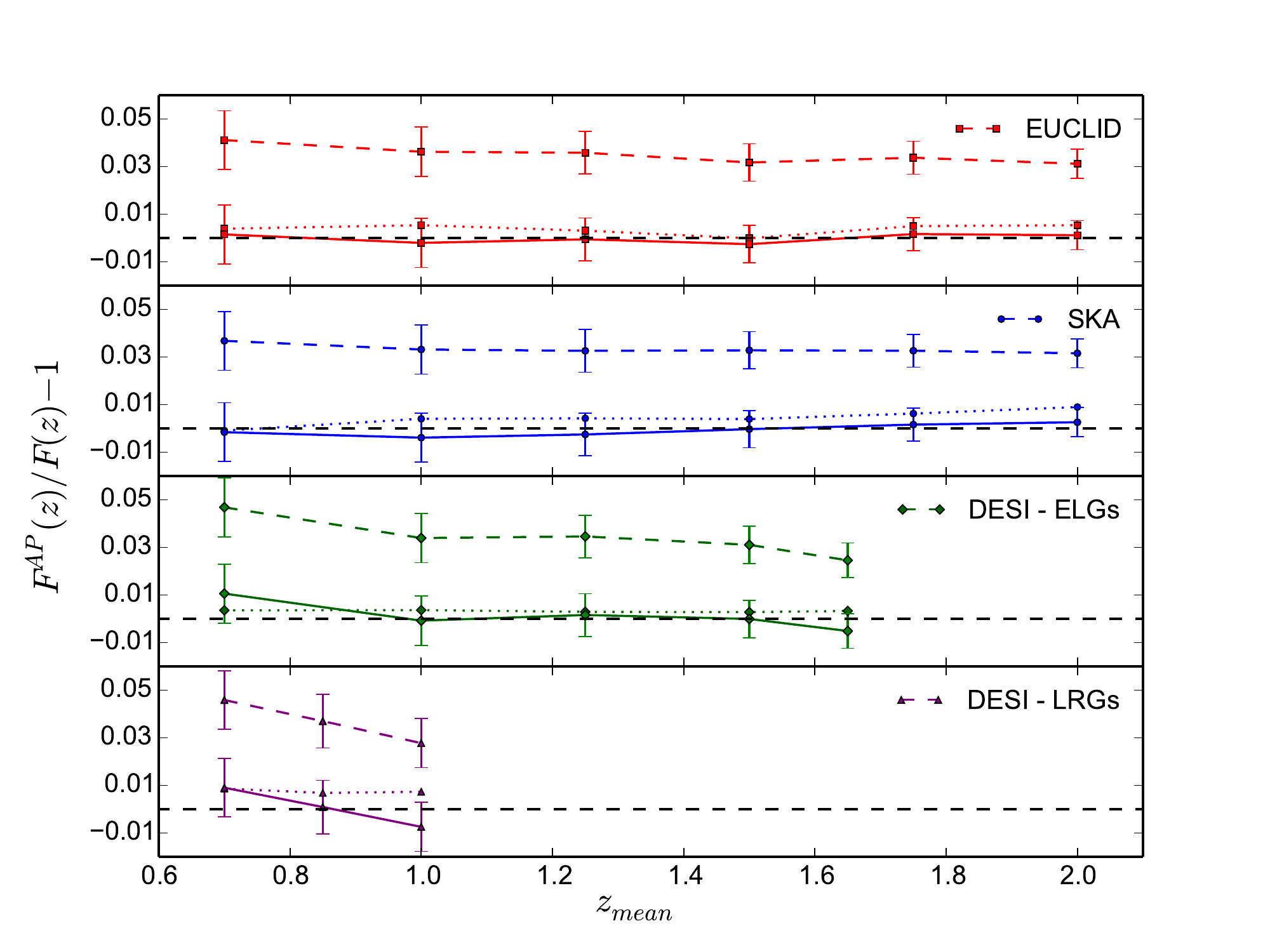}
\caption{The result of the AP test for the four survey specifications considered.
The dash lines refer to the cases in which a window function with
$\sigma_z = 0.02(1+z_{\rm mean})$ is used and no correction is applied in order to take into account the effect of the window function.
The dotted lines refer to the case in which no window function is employed (the small deviations are due to the redshift dependent bias).
The continuous lines represents the cases with window function
in the radial direction and which are corrected using the expression \eqref{fact}, where $\gamma$ are the best-fit values, reported
in Table~\ref{gamma}.}
\label{ap_test_window}
\end{center}
\end{figure}

For all four considered surveys, we
find that the redshift binning generates an offset between $3\%$ and
$5\%$
(see Figure~\ref{ap_test_window}, dash lines).
Minor differences between the four cases are due to the
different galaxy density distributions and the redshift dependence of the galaxy biases.

In order to correct the result of the test from the
distortions induced by the window function, we model the true
AP parameter, as follows from Eqs.~(\ref{correct}, \ref{deltaz_eq}) and as already suggested in~\cite{montanari_durrer}
\begin{equation}
F_{\rm AP}(z_{\rm mean}) = \frac{\Delta z_{\rm BAO}}{\theta_{\rm BAO}}\sqrt{1-\gamma \cdot\Biggl(\frac{\sigma_z}{\Delta z_{\rm BAO}}\Biggr)^2}, \label{fact}
\end{equation}
where $\gamma$ is the parameter which quantifies the 
offset due to the window function.
We estimate the parameter $\gamma$ by minimizing the quantity 
\begin{equation}
\chi^2 =\sum_i\Biggl(\frac{F_{\rm AP}(z_i)-F_{\rm th}(z_i)}{\Delta F_{\rm AP}(z_i)}\Biggr)^2,
\end{equation}
where $F_{\rm th}$ is the expected value for the AP function and 
$\Delta F_{\rm AP}$ is the error on $F_{\rm AP}$, determined by the redshift and angular uncertainties, which we assume to be statistically independent at different redshifts. 
The values of $\gamma$ which minimizes the difference between
the measured AP function and its theoretical values are
reported in Table~\ref{gamma}.

In Figure~\ref{ap_test_window} we show, as solid lines,
the results of the test properly corrected for the window function
effect.
We find that, within the error, the AP function computed
from angular and radial positions of the BAO peak in the correlation
function agrees with the theoretical value. 

\begin{table}[tbp]
\centering
\begin{tabular}{|c|c|c|}
\hline
\multicolumn{1}{|c|}{}                    &
\multicolumn{1}{c|}{Bias}                  &
\multicolumn{1}{c|}{$\gamma$}              \\
Euclid-like     & $\sqrt{1+z}$            & 0.166\\
SKA        & $ c_4\exp{(c_5 z)}$     & 0.161\\
DESI - ELGs & $0.84/D(z)$             & 0.151\\
DESI - LRGs & $1.7/D(z)$              & 0.154\\
\hline
\end{tabular}
\caption{Best-fit values of the parameter $\gamma$, for the
four different survey specifications.
The parameters which specify the bias dependence on redshift for
SKA are taken from Table 4 in \cite{Santos_2014}, telescope SKA (phase 2).
The galaxy bias model for DESI depends on the linear
growth factor $D(z)$.}
\label{gamma}
\end{table}

\subsection{Impact of shot-noise on the AP test}

\label{sec:shotnoise}
In the previous Section, we tested the precision of the AP test for different sources of uncertainties.
In particular, in Section \ref{sec5b} we found that the galaxy bias introduces an extra 1\% error in the estimation of $F_{\rm AP}$.
In this Section we quantify the impact of the shot-noise in the computation of the peak position and on the AP function. We estimate the uncertainty on the cosmological parameters obtained from the AP test including shot-noise and cosmic variance in the error budget. Our estimate is approximate, the true errors are somewhat larger.

We consider an Euclid-like survey, with specifications as described in Appendix~\ref{AP:C}. We first  compute the errors induced by shot-noise and cosmic variance as the statistical 1-$\sigma$ errors on our fitting  parameters for the correlation functions.

The error is computed in two steps. 
First, we include the error due to the shot-noise and cosmic variance in the fitting procedure by minimizing
\begin{equation}
\chi^2 = \sum_i \Biggl(\frac{\xi(\theta_i) - \hat{\xi}(\theta_i, \mathbf{p})}{\sigma_{\xi_{\theta_i}}}\Biggr)^2, \label{chi2_2}
\end{equation}
where $ \hat{\xi}(\theta, \mathbf{p})$ is the parameterization described in Eq.~(\ref{paramet}), $\mathbf{p}$ is the vector of our fitting parameters and  
${\bf \sigma}_{\xi_{\theta_i}}$ are the errors of the correlation function, defined in \eqref{err}. 
Eq. \eqref{chi2_2} implicitly assumes the errors $\sigma_{\xi_{\theta_i}}$ to be uncorrelated, which is not true for the angular correlation function.  Nevertheless, the correlations between errors do not affect the estimated values of the best-fit parameters, but only their covariance. For this reason, we compute the best-fit parameters
by assuming eq. \eqref{chi2_2}  and we separately estimate their errors.
The errors on the parameters $\mathbf{p}$ are estimated as
\begin{equation}
\sigma_{\mathbf{p}} = \sqrt{\mbox{diag}([\mathbf{J^T W J}]^{-1})},
\end{equation}
where $\mathbf{J}$ is the Jacobian of the transformation
\begin{equation}
J^{ij} = \frac{\partial \hat{\xi}(\theta_i, \mathbf{p})}{\partial p_j},
\end{equation}
and $\mathbf{W}$ is the inverse of the full covariance matrix $\mathtt{COV}$,
 \begin{equation}
W^{ij} = [\mathtt{COV}_{\theta^i\theta^j}]^{-1}. 
\end{equation}
%%%%%
The covariance matrix is computed from eq. \eqref{covariance}.

We use a tophat window function of half-width $\sigma_z = 0.02(1+ z_{\rm mean})$, which maximizes the signal-to-noise ratio as 
shown in Section~\ref{window}.
%%%%%%%%%%%%%
\begin{table}[tbp]
\centering
\begin{tabular}{|c|cc|cc|}
\hline
\multicolumn{1}{|c|}{}                    & 
 \multicolumn{2}{c|}{$\sigma_{\theta_{\rm BAO}}$ [\%]}&
  \multicolumn{2}{c|}{$\sigma_{\theta_{\rm BAO}}$ [\%]} \\
\multicolumn{1}{|c|}{}                    & 
 \multicolumn{2}{c|}{Cosmic variance + shot-noise}&
  \multicolumn{2}{c|}{Cosmic variance} \\
\hline
\multicolumn{1}{|c|}{z}                         &
\multicolumn{1}{c}{Polynomial}                &
\multicolumn{1}{c|}{Power-law}               & 
\multicolumn{1}{c}{Polynomial}                &
\multicolumn{1}{c|}{Power-law}               \\
\hline
0.7  &    2\%   &  1.4\%  &    2\%   &  1.4\%     \\
1.0  &    3\%   &  1.9\%  & 1.7\%   &  1.0\%     \\
1.25&    3\%   &  1.9\%  & 1.6\%   &  0.9  \\
1.5  &    3\%   &   2.3\%  & 1.5\%  &  0.8  \\
1.75&    5\%   &   3\%     & 1.4\%  &  0.9\%  \\
2.0  &  12\%   &   8\%     & 1.3\%  & 0.7\%  \\
\hline
\end{tabular}
\caption{\label{tab_shotnoise} 
1-$\sigma$ statistical error on $\theta_{\rm BAO}$ estimated by the fit for two different parameterizations.
The polynomial parameterization is the \emph{polynomial+Gaussian} we adopted throughout the paper, where the
polynomial degree is set to be $N=10$, while the power-law is the 6-parameter \emph{power-law+Gaussian} model
considered in Appendix \ref{Ap:B}.
}
\label{table_param}
\end{table}

In Table~\ref{tab_shotnoise} we report the errors on $\theta_{\rm BAO}$ computed at different redshifts for two different parameterizations. The polynomial model is the $10$ degrees \emph{polynomial+Gaussian} we tested in the previous sections, see Eq.~\eqref{paramet},
which fits 14 parameters. The \emph{power-law+ Gaussian} model refers the parameterization tested in 
 Appendix \ref{Ap:B}, see Eqs.~(\ref{xiperp_power}, \ref{exp}), which is based on 6 parameters.
In both cases, we find that the statistical errors 
dominate the uncertainties due to the angular resolution of the survey.  
 At low redshift ($z < 1.5$), the statistical error is dominated by cosmic variance, while shot-noise dominates at high redshifts ($z \ge 1.5$), where the error rapidly increases up to $\sim 10\%$ (see Table \ref{tab_shotnoise}).
Somewhat surprisingly, the errors for the \emph{power-law + Gaussian} model
are smaller than the ones estimated for the \emph{polynomial+Gaussian} model, as shown in Table \ref{tab_shotnoise}.  

This suggests that a fitting model with 14 parameters, even if more accurate, requires more data to constrain simultaneously the free parameters and the cosmological model. 
Table ~\ref{tab_shotnoise} shows that a power-law still gives better constraints.
Nevertheless, the power-law template may introduce a bias in the result, as it is shown in
Appendix \ref{Ap:B}. Therefore, it can be worth to partially sacrifice the precision of the result 
in favor of a better accuracy. 

In order to estimate the impact of these errors on the AP function we compute the
total error by summing the systematic errors, due to the resolution of the survey and on the bias,
and the statistical errors, due to the shot-noise and cosmic variance,
\begin{equation}
\frac{\Delta F_{\rm AP}}{F_{\rm AP}} = \sqrt{ \Biggl( \frac{\Delta F_{\rm AP}}{F_{\rm AP}}\Biggr)_{\rm sys}^2+ \Biggl(\frac{\Delta F_{\rm AP}}{F_{\rm AP}}\Biggr)_{\rm stat}^2}.
 \end{equation}
%%%%%%%%%%%%%%%
\begin{figure}[t!]
\begin{center}
\includegraphics[width=0.8\textwidth]{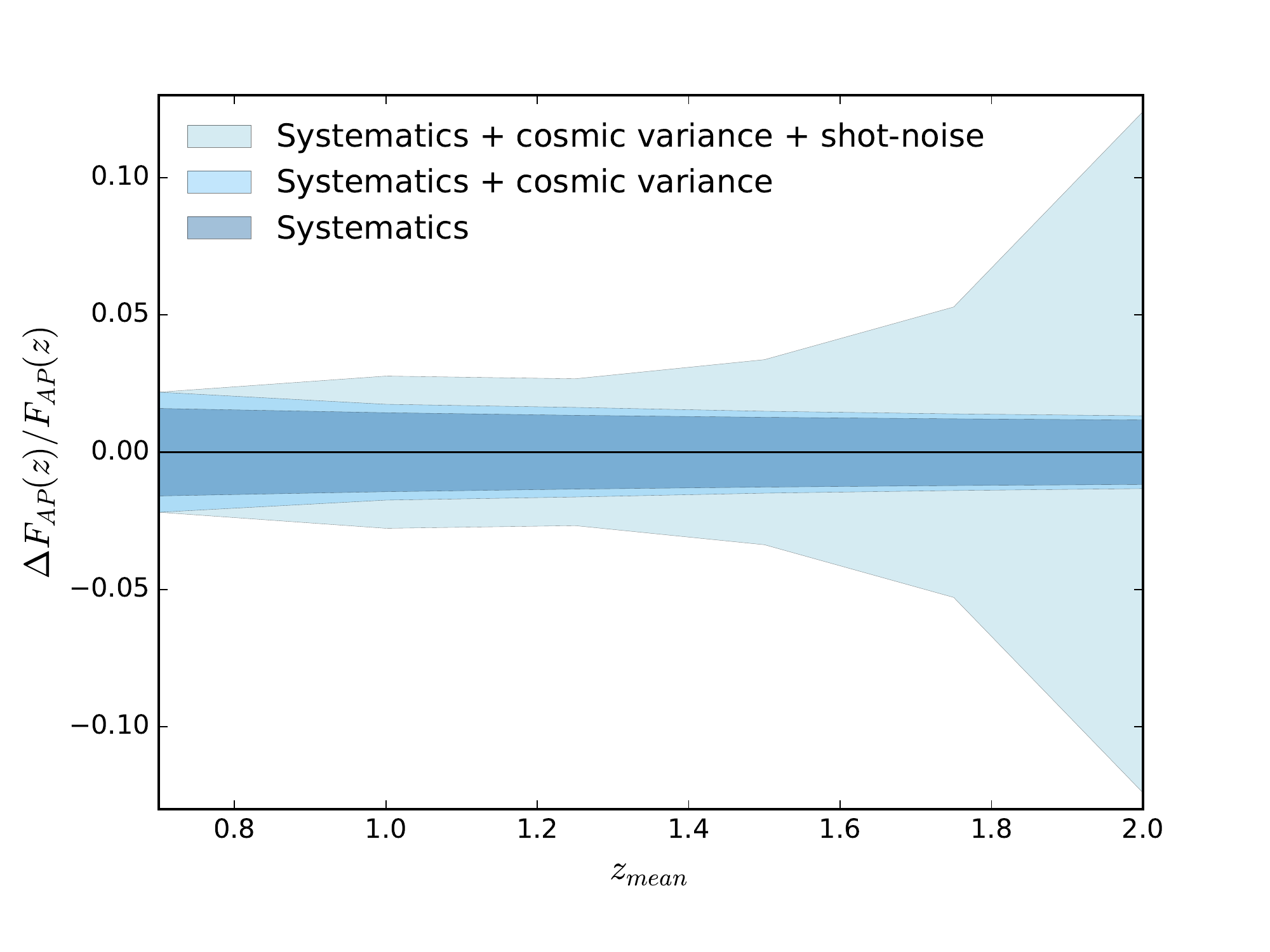}
\caption{Impact of shot-noise on the error of the AP function. }
\label{F_err}
\end{center}
\end{figure}
%%%%%%%%%%%%
In Figure \ref{F_err} we show the impact of the error due to the shot-noise on the AP function.
When only the systematic error is taken into account, the error is dominated by
the resolution in redshift in the radial correlation function. 
The shot-noise contribution is, however, not negligible
and becomes even dominant at high redshifts, due to the lower galaxy density. This indicates that with a higher number of galaxies, one could improve the AP test significantly.
 If both  sources of uncertainty are taken into account we find a relative error between 3\% and 4\% at $z \le 1.5$, while at higher redshift the error  increases up to 12\% at $z=2$.

In the low redshift regime ($z \le 1.5$), the estimated errors for the AP parameter $F_{\rm AP}$
can be compared to the recent analysis presented by the Baryon Oscillation Spectroscopic Survey (BOSS) collaboration \cite{BOSS2016}. 
In \cite{BOSS2016} the AP parameter is estimated in three redshift bins in the range
$z \in [0.2, 0.75]$ and in each bin the analysis uses the full two-dimensional correlation function.
For all the three bins, the errors on $F_{\rm AP}$ amount approximatively to $3\%$ (see Table 4 in
\cite{BOSS2016}), which is the same order of the errors in Figure \ref{F_err} in the low redshift regime.

\subsection{Constraints on cosmological parameters}
To conclude,  we give  in this section an approximative estimation of the expected errors on $\Omega_{\rm m}$, $\Omega_{\rm k}$
and $w_{\rm DE}$ from the AP test. 

\begin{figure}[t!]
    \begin{subfigure}[b]{0.5\textwidth}
        \centering
        \includegraphics[width=\textwidth]{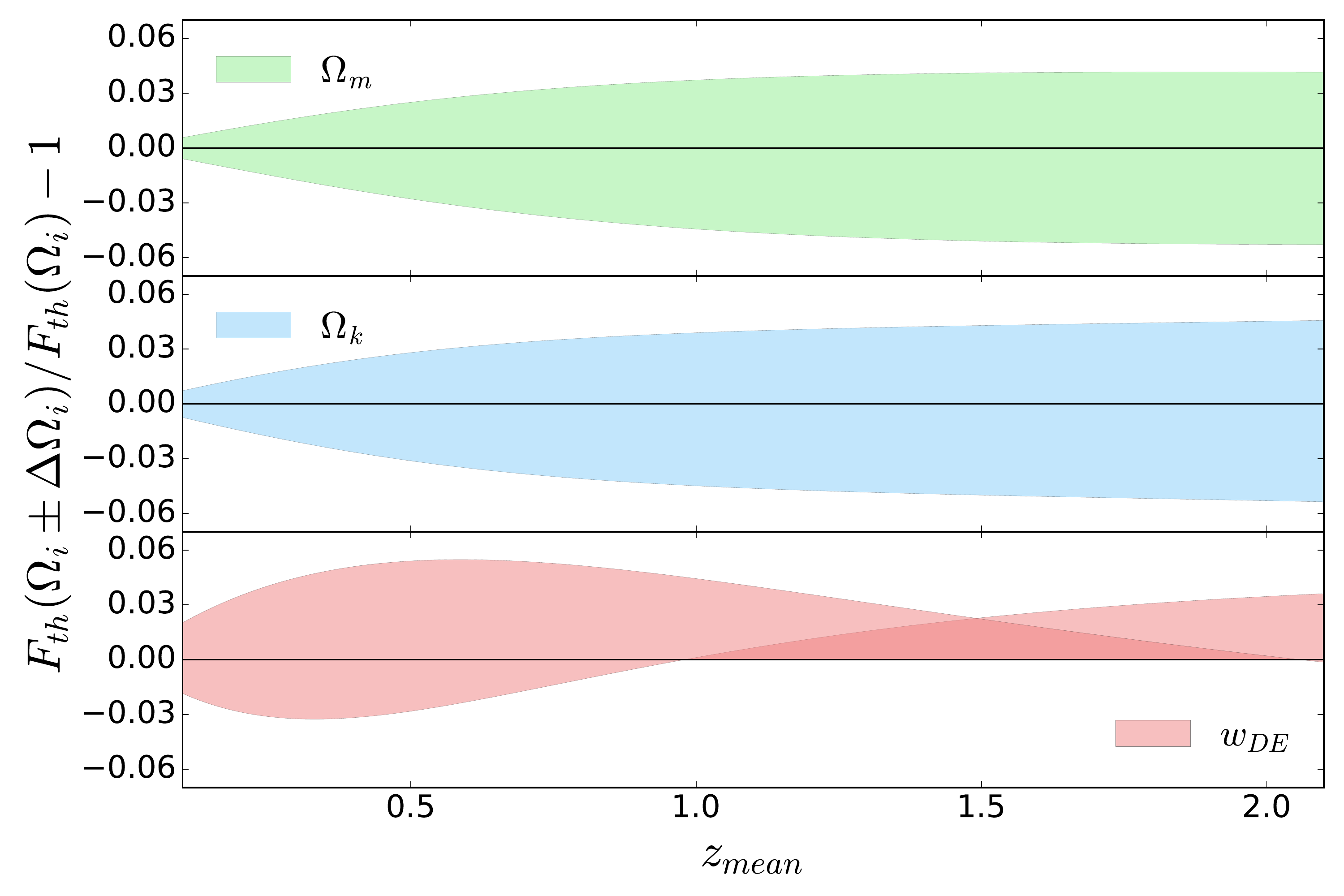}
         \caption{BOSS - Galaxy clustering only}
        \label{boss}
    \end{subfigure}
    \begin{subfigure}[b]{0.5\textwidth}
        \includegraphics[width=\textwidth]{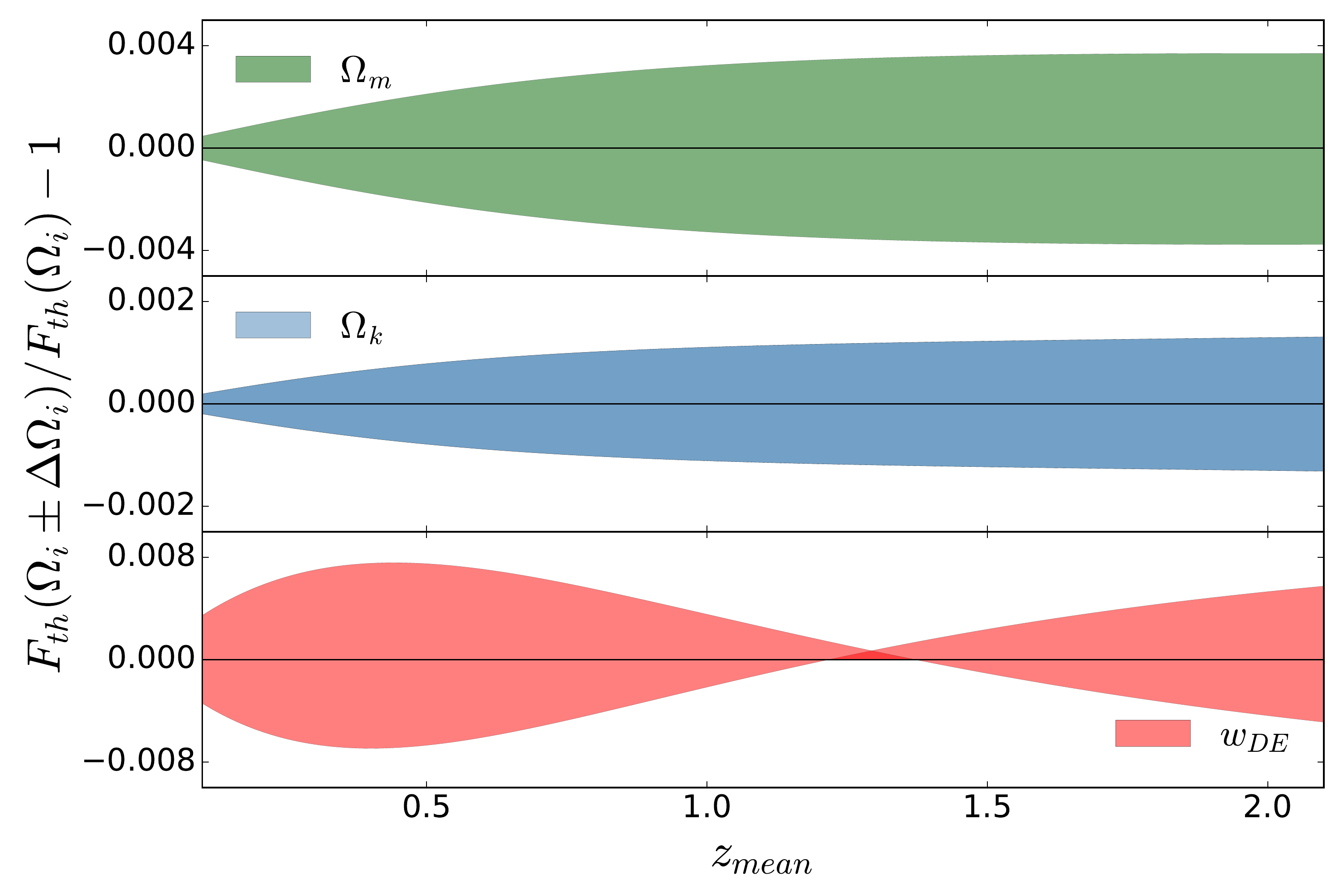}
     \caption{Planck 2015}
    \label{planck}
    \end{subfigure}
\caption{Sensitivity of the AP function to different cosmological parameters. $\Delta \Omega_i$ is taken to be the error on the parameter $\Omega_i$ from galaxy clustering only (top panel) and
from CMB analysis (bottom panel).
The constraints from galaxy clustering come from the Baryon Oscillation Spectroscopic Survey (BOSS) of SDSS-III, when only the CMASS-Large dataset is employed \cite{Chuang}. The constraints from CMB are the 1-$\sigma$ limits ($\Omega_{\rm m}$) and 2-$\sigma$ limits ($\Omega_{\rm k}$ and $w_{\rm DE}$) in Planck 2015 analysis \cite{planck2015}, when \emph{CMB plus external data} are used. The sign change in $\frac{\partial F_{\rm AP}}{\partial w_{\rm DE}}$ at $z\simeq 1.3$ is well visible in the precise Planck data  'smeared out' in the less accurate BOSS data.}
\label{fig_DeltaF}
\end{figure}

First we study the dependence of the AP function on cosmological parameters.
In Figure \ref{fig_DeltaF} we show the sensitivity of the AP function to variations of the cosmological parameters when considering BOSS error-bars from only galaxy clustering  \cite{Chuang} (left panel), and the Planck 2015 error-bars \cite{planck2015}(right panel).

For each parameter, we also naively compute its change over the allowed range of variation of the AP function when all the other parameters are kept fixed,
\begin{equation} 
\Delta \Omega_i(z) = \Biggl(\frac{\partial F_{\rm AP}}{\partial \Omega_i}\Biggr)^{-1} \Delta F_{\rm AP}. \label{error_1red}
\end{equation}
We use for $\Delta F_{\rm AP}$ the error computed in Section \ref{sec:shotnoise} and plotted in Figure \ref{F_err}.

\begin{figure}[t!]
\begin{center}
\includegraphics[width=0.8\textwidth]{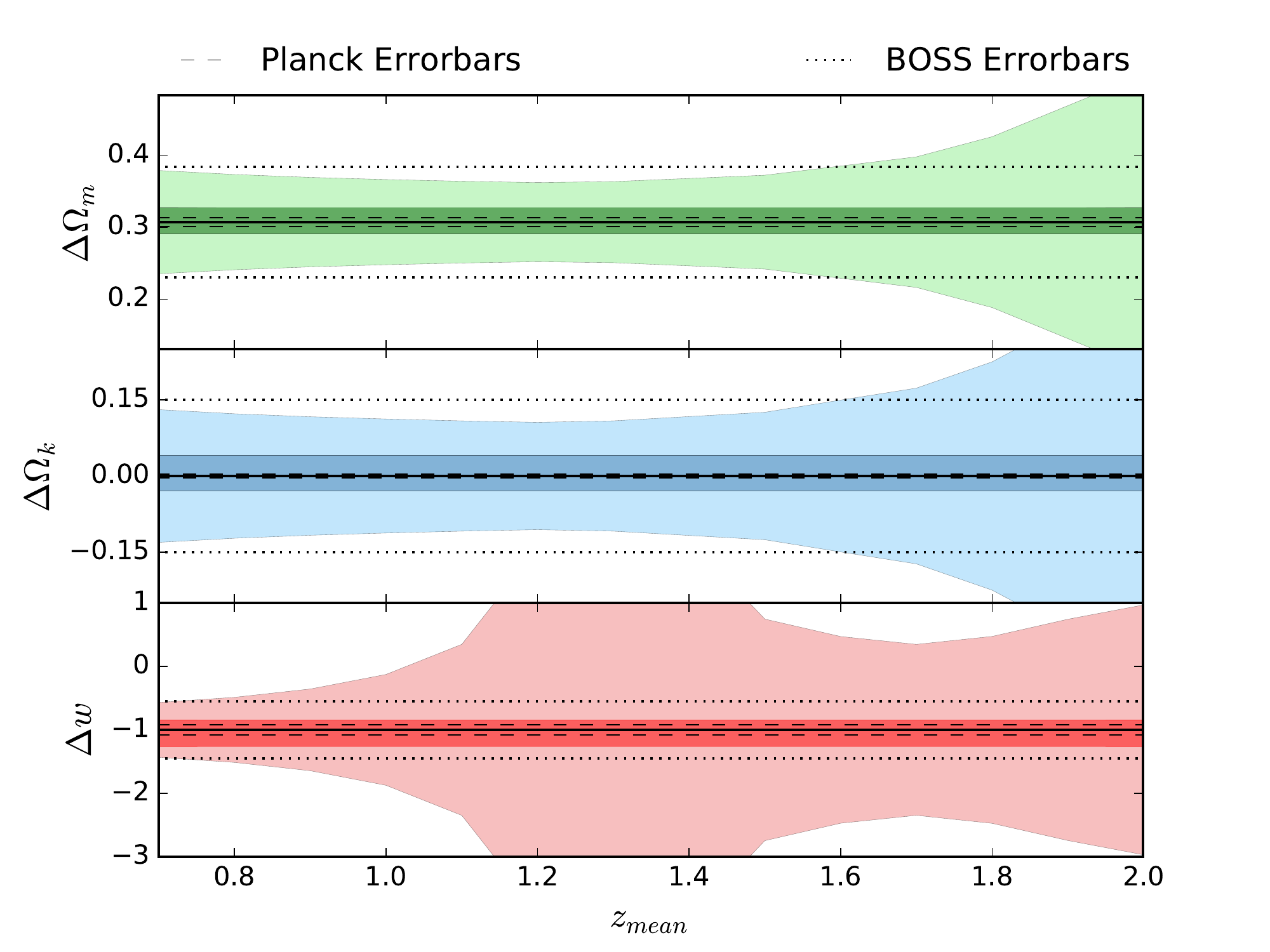}
\caption{Constraints on cosmological parameters from a single measurement of the AP function (light shaded region) and from $N=14$ measurements (assumed to be independent) (dark shaded region). The Planck and BOSS limits are indicated by dashes and dotted horizontal lines.}
\label{dOmega}
\end{center}
\end{figure}

In Figure \ref{dOmega} we compare $\Delta \Omega_i(z)$ to the constraints presently available  from galaxy clustering data, more precisely from BOSS \cite{Chuang}, and from the state-of-the-art constraints,  given by Planck, \emph{CMB + External Data} analysis \cite{planck2015}.
The errors  $\Delta \Omega_i(z)$ are represented by the light shaded regions, while 
the errors from BOSS (only galaxy clustering) and Planck 2015 are shown in dotted and dashed line, respectively. 
%%%%%%%%%%%%
%%%%%%%%%%%%

The equation of state of dark energy, for this simple parameterization, is poorly constrained in the AP test around $z \sim 1.3$. This reflects the fact that the AP parameter is not sensitive to $w$ around $z \sim1.3$, as we see in Figure~\ref{fig_DeltaF}.

The light shaded regions in Figure \ref{dOmega} represent the error on a single measurement of $F_{\rm AP}$, the overall statistical errors on the parameters
can be reduced by performing $N$ independent measurements at different redshifts.
%%%%%%%%%%%%%

We also estimate the errors from $N = 14$ independent measurements 
in the range $[0.7, 2.0]$ with redshift separation $\De z=0.1$. 
The errors are estimated by fitting
a set of hypothetical data points for the AP function $F_{\rm DATA}$. The data points are computed by summing 
the value of $F_{\rm AP}$ for our fiducial cosmology to a randomly generated scattering term proportional
to the error.
The fit is performed by
running $\mathtt{emcee}$ \cite{emcee}, a Markov Chain Monte Carlo (MCMC) code.
The $\chi^2$ inferred from the simulated data  is 
\begin{equation}
\chi^2 = \sum_{i=1}^N \Biggl(\frac{F_{\rm DATA}(z_i) - F_{\rm AP}(z, \Omega) }{\Delta F_{\rm AP}}\Biggr)^2,
\end{equation} 
where we vary only one parameter $\Omega$ at a time, keeping the others fixed to their fiducial value.

We obtain the following results for $\Omega_m$, $\Omega_k$ and $w_{\rm DE}$:
\begin{equation}\label{eq:err14}
\Omega_m = 0.309^{+0.019}_{-0.018}, \qquad  \Omega_k = 0.005^{+0.036}_{-0.034},
\qquad w_{\rm DE} = -1.02^{+0.18}_{-0.24},
\end{equation}
where the quoted errors are the $1-$sigma limits.

The dark shaded regions in  Fig.~\ref{dOmega} shows the error regions given in (\ref{eq:err14}).
The error-bars from the AP test are still significantly larger than the Planck error bars but smaller that present errors obtained from galaxy clustering only.

The most recent analysis combining CMB data with BAO and redshift space distortions
is presented by the BOSS collaboration \cite{BOSS2016_full}.
The errorbars on the three cosmological parameters $\Omega_m$, $\Omega_k$ and $w_{\rm DE}$
are comparable to the Planck 2015 errorbars and therefore result to be significantly
smaller than the errors quoted in \eqref{eq:err14}.
This is not surprising since the AP test uses just one single scale in the power spectrum, while parameter estimation from the full CMB power spectrum (Planck), or the full shape of the correlation function (BOSS), has the advantage of using all available  modes.

\section{Conclusions}
\label{conc}
We have presented a model independent method to perform the Alcock Paczy\'nski  test. We applied the test to the BAO feature in the galaxy 2-point correlation function, in the radial and transverse directions. In order to perform the test without  prior assumption on the cosmological parameters, we model the shape of the correlation function with a polynomial, while the acoustic feature is modeled as a Gaussian. In this work we have shown that both redshift space distortions and galaxy bias must be taken into account
to improve the accuracy of the estimated BAO scale, while gravitational lensing does not significantly affect the result of the AP test. Finally, we have investigated the projection effect induced by the finite width of the radial window function. In fact, in the transverse direction a redshift bin of a typical width of a photometric survey is sufficient to locate the peak position and it can be employed to maximize the signal-to-noise. However, the projection effect shifts the measured acoustic peak toward smaller angular scale. If not corrected for, this  introduces a systematic error of up to $5\%$ in the AP test. We have computed the correction that has to be applied in order to adjust the result of the test for this effect for three  planned galaxy surveys, Euclid-like, SKA and DESI. We have also estimated the effect of shot-noise and cosmic variance in the precision of the BAO peak detection and on the AP test. We have found that the precision of the test could be improved significantly by increasing the number of galaxies in the survey, especially at high redshift. Finally, we have shown that cosmological parameters can be estimated in a model independent approach, by the AP test, with roughly the same accuracy as the BAO analysis performed with the BOSS survey. The latter, however, has the disadvantage to use the power spectrum in Fourier space, $P(k)$, which itself depends in a non-trivial way on cosmological parameters via the conversion of angles and redshifts into distances.

\acknowledgments
We thank David Alonso and Francesco Montanari for useful discussions.
MV and ED are supported by the cosmoIGM starting grant. FL, CB, MV and ED are supported by the INFN PD51 grant INDARK.
RD thanks the Swiss National Science Foundation for financial support. 
CB, MV acknowledge support by the Italian Space Agency through the ASI contracts Euclid-IC (I/031/10/0). 
We thank the Galileo Galilei Institute for Theoretical Physics for the hospitality and the INFN for partial support during the completion of this work.
\vspace{1.5cm}

\appendix

\noindent{\LARGE \bf Appendix}

\section{Methodology tests}
\label{Ap:B}

Here we report the results of some tests we performed in order
to find the best method to model the transversal and longitudinal correlation functions and in order to verify its reliability.
 %%%%%%%%%%%%
\begin{figure}[t!]
\begin{center}
\includegraphics[width=0.8\textwidth]{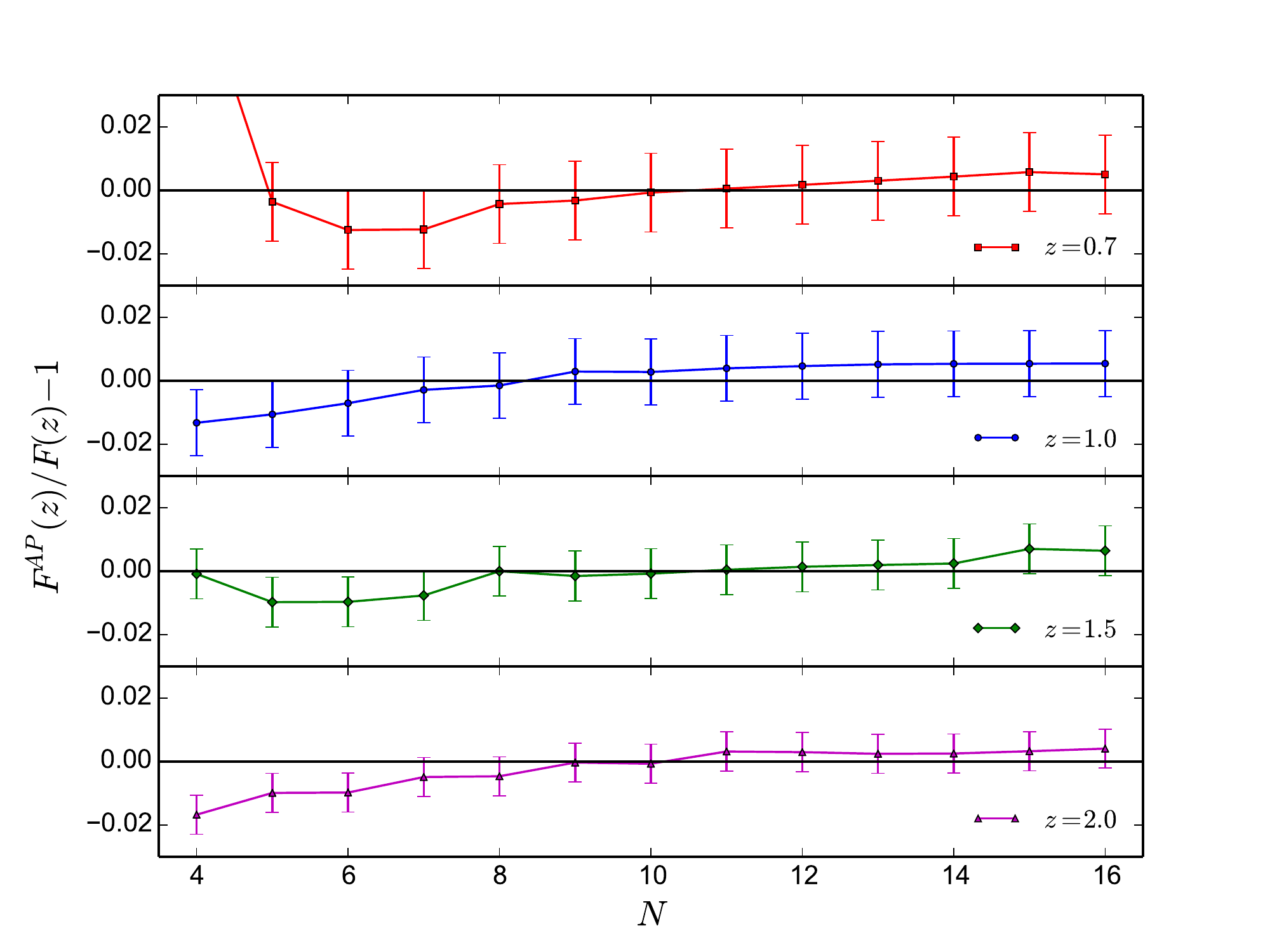}
\caption{AP consistency test for different degrees $N$ of the polynomial
in the parameterization of the correlation function.
The error-bars are computed from linear propagation of uncertainty, eq. \eqref{prop_err}, assuming
the uncertainty on $\Delta z_{BAO}$ and $\theta_{BAO}$ to be the resolution of the galaxy survey.}
\label{Poly_test}
\end{center}
\end{figure}
  %%%%%%
We test our template model, Eq.~\eqref{paramet}, for different degrees of the fitting polynomial.
Figure~\ref{Poly_test} shows a consistency check for 
values of $N$ in the range $[4:16]$. 
The correlation functions used in this test are computed including
the local density term and the redshift space distortions correction.

The errorbars are computed from linear propagation of uncertainty,   
\begin{equation}
\frac{\Delta F_{\rm AP}}{F_{\rm AP}} = \sqrt{\Biggl(\frac{\delta z}{\Delta z_{\rm BAO}}\Biggr)^2 + \Biggl(\frac{\delta \theta_{\rm BAO}}{\theta_{\rm BAO}}\Biggr)^2}, \label{prop_err}
\end{equation}

where $\delta z = 6.25 \cdot 10^{-4}$ is the resolution used for the radial correlation function, and $\delta \theta_{\rm BAO} = 0.1$ arcsec is the angular resolution of an Euclid-like survey.

For these values, the uncertainty due to the angular resolution can be safely neglected
and the relative error can be written as
\begin{equation}
\frac{\Delta F_{\rm AP}}{F_{\rm AP}} = \frac{\delta z}{\Delta z_{\rm BAO}}.
\end{equation}

Since $\delta z$ is chosen to be constant, the errorbars decrease at larger redshift
because the radial BAO peak position corresponds to larger $\Delta z_{\rm BAO}$.

%They decrease with redshift because we are using are using a fixed resolution at different redshift
%($\delta (\Delta z_{BAO}) = \mbox{constant}$), while the BAO peak position in the radial correlation
%function, $\Delta z_{BAO}$, increases with redshift.

\begin{figure}[!ht]
\begin{center}
\includegraphics[width=0.8\textwidth]{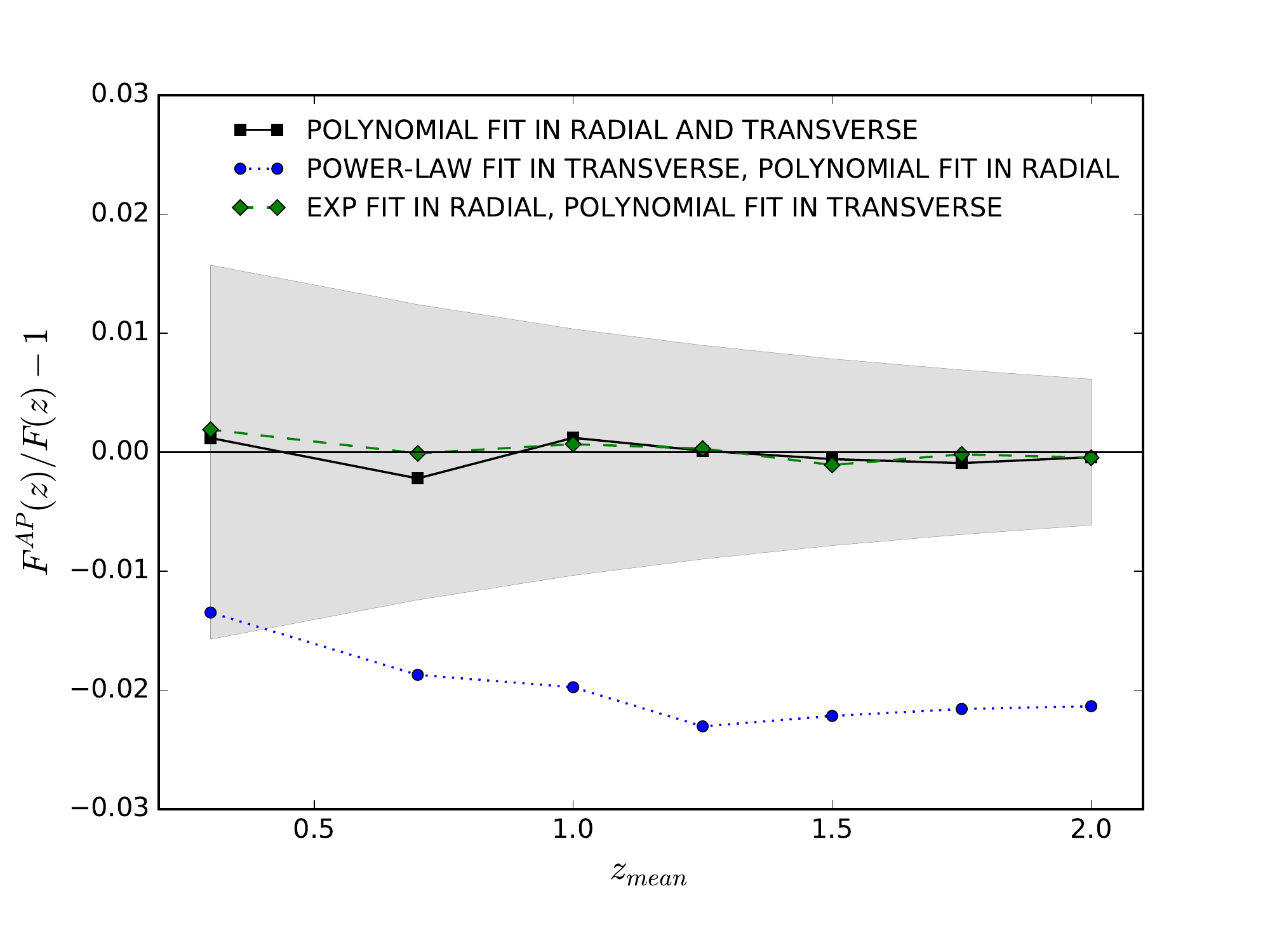}
\caption{AP consistency test at different redshifts, for different parameterization of the
radial and transverse correlation functions. The continuous black line
refers to the test performed using a polynomial fitting function for 
the estimation of both $\Delta z_{\rm BAO}$ and $\theta_{\rm BAO}$, the
dot blue line is computed using a polynomial fit in the longitudinal direction and the \emph{power-law+Gaussian} model in the transverse one, while the dash green line is computing using a polynomial for
$\xi_{\perp}(\theta)$ and the exponential fit in \eqref{exp} for
$\xi_{\parallel}(\Delta z)$. }
\label{param}
\end{center}
\end{figure}

We found that for $N$ between 8 and 14 the relative difference
between the fitted AP parameter and its theoretical value is smaller
than $0.5 \%$ at all redshifts. 
We checked the accuracy of the fit comparing the sum of the squared deviations between our fit and the {\sc class}gal output and we found the  most numerically accurate results satisfying the AP test. Therefore, we choose a polynomial of degree $N = 10$  to perform the next analysis.
In order to test the goodness of this parameterization, we perform
the same analysis for different cosmologies, including a cosmology
with a dynamical dark energy equation of state. 
In all cases we find qualitatively the same behavior shown in Figure~\ref{Poly_test}.
We compare our parameterization also with others
described in the literature.
In particular, the transverse correlation function is often
modeled by a \emph{power-law + Gaussian} function~\cite{Sanchez2}
\begin{equation} \label{xiperp_power}
\xi_{\perp}(\theta) = A + B \cdot \theta^{\gamma} + C \cdot e^{-(\theta-\theta_{\rm BAO})^2/2\sigma^2},
\end{equation} 
with $A$, $B$, $\gamma$, $C$, $\theta_{\rm BAO}$ and $\sigma$
as free parameters.
As for the radial correlation function, we compare our model
with the parameterization used in~\cite{Sanchez1}
\begin{equation}
\xi_{\parallel}(\Delta z) = A + B  \cdot e^{-C\Delta z} - D  \cdot e^{-E \Delta z}
+ F \cdot e^{-(\Delta z-\Delta z_{\rm BAO})^2/2\sigma^2}, \label{exp}
\end{equation} 
where the free parameters are $A$, $B$, $C$, $D$, $E$, $F$, $\Delta z_{\rm BAO}$  and $\sigma$.
The result of this consistency test is shown in Figure~\ref{param}.
We find that both the exponential parameterization, Eq.~\eqref{exp},
and the 10 degree polynomial, Eq.~\eqref{paramet}, fit well the radial correlation function
(differences are smaller than 0.5\% at all redshifts for the AP consistency check). 
On the other hand, we find a discrepancy between the
power-law and the polynomial parameterization for the transverse
correlation function.
%%%%%%%%%%%%
%%%%%%%%%%
In Table~\ref{table_param} we compare the  comoving BAO scale
computed from the peak positions in the transverse and radial
direction.

%\FRArem{We find that} {\Myg On the other hand}, using a \emph{power-law+Gaussian} fit 
%for the transverse correlation function, the peak position
%is systematically shifted towards larger scales. This systematic
%introduces a 2\% offset in the AP test.
%{\Myg The \emph{power-law + gaussian} model for the transverse correlation function was designed for photometric survey and it was tested with N-body simulations.
%In this case the uncertainty due to the method is estimated to amount to 1\%.
%In this main source of systematic is due to the projection effect induced by the radial
%window function. In this case the }
We find that, using a \emph{power-law+Gaussian} fit 
for the transverse correlation function, the peak position
is systematically shifted towards larger scales. This systematic
introduces a 2\% offset in the AP test.

The polynomial and the exponential parameterization
gives consistent values of the BAO scale at all redshifts.
However, the polynomial fit is
able to model the data over a larger range, therefore we adopt this parameterization.

\begin{table}[tbp]
\centering
\begin{tabular}{|c|cc|cc|}
\hline
\multicolumn{1}{|c|}{$z$}                    & 
\multicolumn{2}{|c|}{$L_{\perp} (Mpc/h)$}      &
 \multicolumn{2}{|c|}{$L_{\parallel} (Mpc/h)$} \\
 \hline
\multicolumn{1}{|c|}{}                        &
\multicolumn{1}{c}{Polynomial}                &
\multicolumn{1}{c}{Power-law}                 &
\multicolumn{1}{|c}{Polynomial}               &
\multicolumn{1}{c|}{Exponential}               \\
\hline
0.3 &  101.7 $\pm$ 0.2  & 103.2 $\pm$  0.2 & 101.9 $\pm$ 1.6  & 102.0 $\pm$ 1.6\\
0.7 &  101.9 $\pm$ 0.3  & 103.6 $\pm$  0.3 & 101.7 $\pm$ 1.3 &  101.9 $\pm$ 1.3\\
1.0 &  101.8 $\pm$ 0.4  & 103.9 $\pm$  0.4 & 101.9 $\pm$ 1.1 &  101.8 $\pm$ 1.1\\
1.25& 101.8 $\pm$ 0.5 &  104.3 $\pm$  0.5 &  101.9 $\pm$ 0.9 & 101.9 $\pm$ 0.9\\
1.5 &  101.9 $\pm$ 0.5 &  104.1 $\pm$  0.5 &  101.8 $\pm$ 0.8 & 101.8 $\pm$ 0.8\\
1.75& 101.9 $\pm$ 0.6 &  104.0 $\pm$  0.6 &  101.8 $\pm$ 0.7 & 101.8 $\pm$ 0.7\\
2.0 &  101.9 $\pm$ 0.6 &  104.0 $\pm$  0.6 &  101.8 $\pm$ 0.6 & 101.8 $\pm$ 0.6\\
\hline
\end{tabular}
\caption{\label{tab:i} The comoving BAO  scale in units of Mpc$/h$ computed using  the different parameterizations given in the text for the correlation function at different redshifts.
}
\label{table_param}
\end{table}

\section{Surveys specifications} 
\label{AP:C}

In this section we report the specifications used
for the different surveys considered in 
Section~\ref{window}.
The parameters of the surveys are
the sky coverage $f_{\rm sky}$, the number of galaxies per unit redshift per
square degree and the redshift dependence of the galaxy bias.

\begin{figure}[t!]
    \begin{subfigure}[b]{0.5\textwidth}
        \centering
        \includegraphics[width=\textwidth]{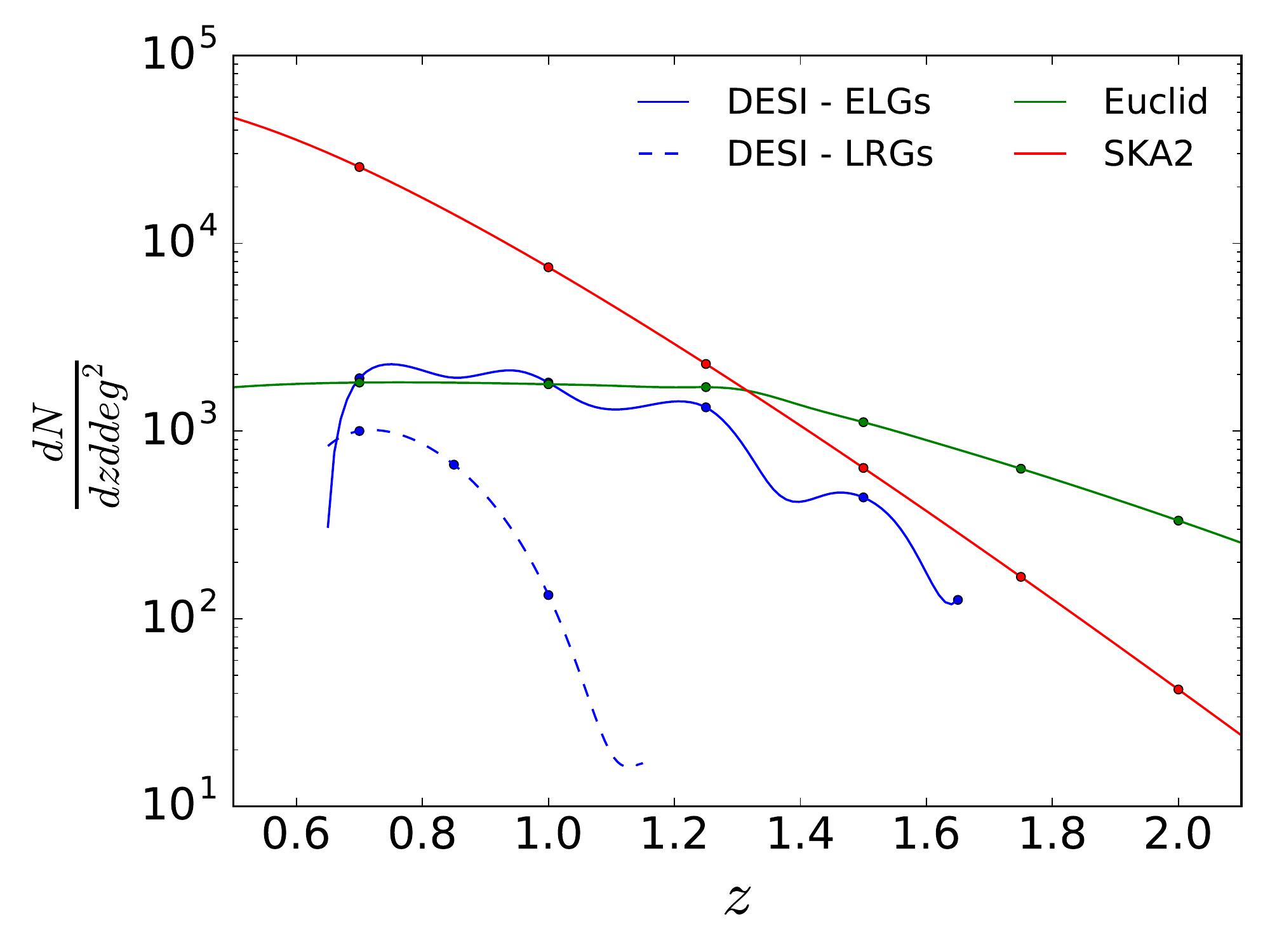}
     \caption{Number of galaxies per square degree per unit redshift.
     }   
     \label{dndz}
    \end{subfigure}
    \begin{subfigure}[b]{0.5\textwidth}
        \includegraphics[width=\textwidth]{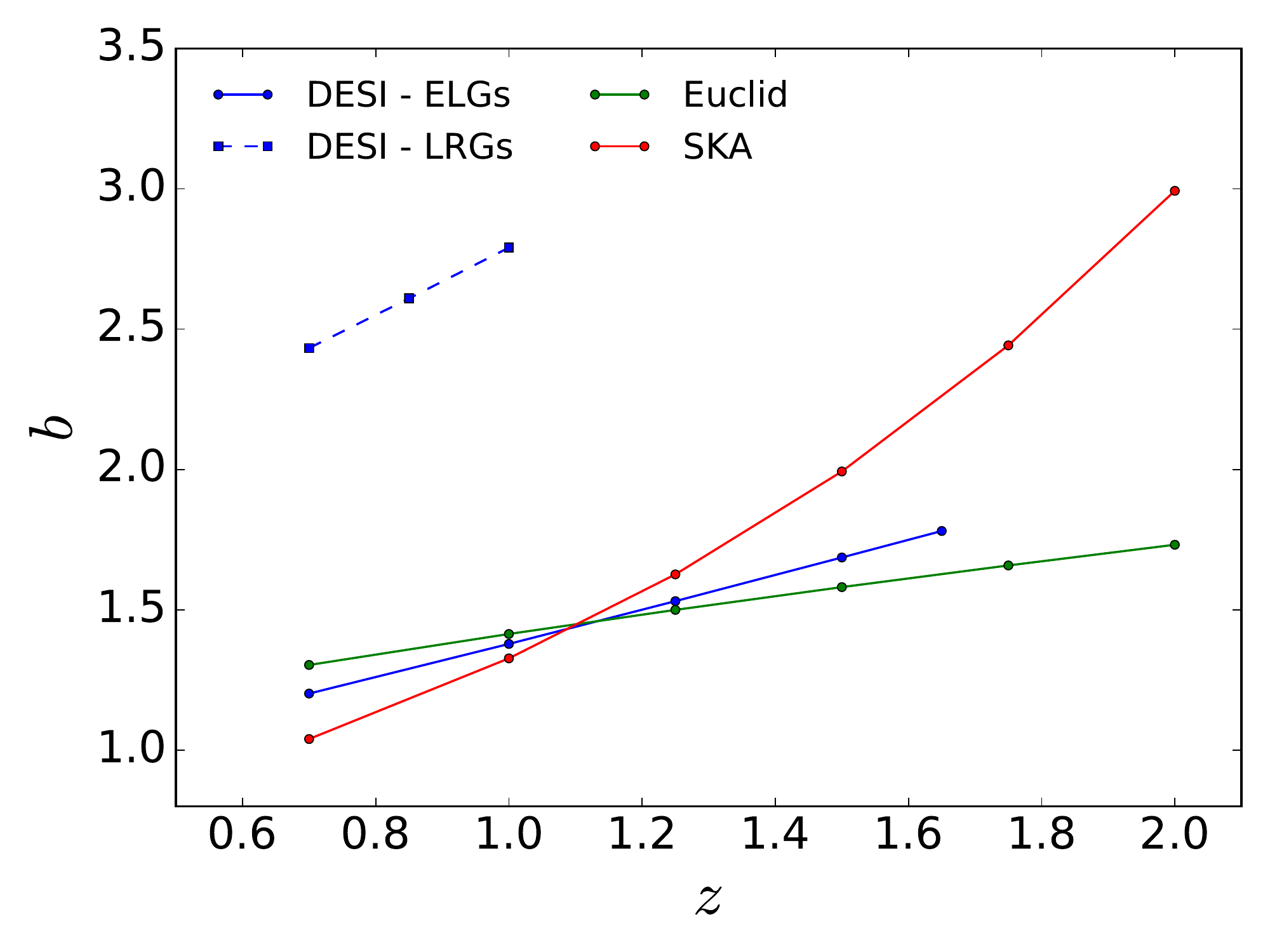}
    \caption{\\Redshift evolution of the galaxy bias.}
    \label{b}
    \end{subfigure}
\caption{Figure~\subref{dndz} and Figure~\subref{b} show, respectively,
the number of galaxies per square degree per unit redshift and the
redshift dependent bias for the four cases described in this Appendix:
Euclid, SKA, DESI ELGs (bright Emission Line Galaxies) and DESI LRGs (Luminous Red Galaxies).
The markers identify the redshifts for which we performed the AP test in Section~\ref{window}.
}
\label{survey}
\end{figure}
  
\subsection{Euclid} 
Our Euclid-like survey is modeled
following  Appendix A.3 in~\cite{Audren}.
We assume a sky fraction $f_{\rm sky} = 0.375$.
The number of galaxies per unit redshift per square degree
is computed from Table 2 of~\cite{Geach}, for the case of a limiting flux
of $3 \times 10^{-16} \rm {erg}\, \rm{s}^{-1}\, \rm{cm}^{-2}$.
Following Ref.~\cite{Audren}, we multiply the tabulated values by an efficiency factor $0.25$ and we divide them by the factor $1.37$ to
get conservative prediction.
The redshift dependence of the bias is modeled as in the forecasts
presented in~\cite{Amendola}
\begin{equation}
b(z) = \sqrt{1+z}.
\end{equation}
\subsection{SKA} 
We used the technical specification for SKA
reported in~\cite{Abdalla}.
Galaxy number density per unit redshift per square
degree and bias evolution are given by 
\begin{align}
\frac{dN}{dzd\Omega} &= 10^{c_1} z^{c_2} \exp{(-c_3 z)},\\
b(z) &= c_4 \exp{(c_5 z)},
\end{align}
where we used for the coefficients $c_i$ the best-fit values 
reported in Table 4 of~\cite{Santos_2014}, for SKA2. 
For SKA2, the sky coverage is expected to be around $30000 \, \mbox{deg}^2$,
which corresponds to a sky fraction $f_{\rm sky} = 0.727$.
\subsection{DESI}

Survey specifications for DESI, for both
ELGs and LRGs, are taken from the 
Science Technical Design Report~\cite{desi}.
The number of galaxies per unit redshift per square degree,
for both ELGs and LRGs, are assumed to be the one
reported in Table 2.3 of~\cite{desi}.
The values at intermediate redshifts are computed by interpolation.
The redshift dependence of galaxy bias is expressed in terms
of the linear growth factor $D(z)$.
For the ELGs we have
\begin{equation}
b(z) = \frac{0.84}{D(z)},
\end{equation}
while for LRGs we assume
\begin{equation}
b(z) = \frac{1.7}{D(z)}.
\end{equation}
Following~\cite{desi}, we assume for DESI a sky coverage equal
to $14000\,\mbox{deg}^2$, which corresponds to a sky fraction
$f_{\rm sky} = 0.339$. 
In Figure~\ref{survey} we compare the expected number of galaxies
per unit redshift, per square degree and the bias evolution
for the 3 cases discussed above.
The number of galaxies per unit redshift per square degree 
is relevant for the computation of  shot noise.

The number of galaxies per steradian $n_i$ inside the $i$-th redshift bin is computed as
\begin{equation}
n_i = \frac{1}{4\pi}\int d\Omega \int_{z_{\rm mean} - \Delta z/2}^{z_{\rm mean}+ \Delta z/2}
\frac{dN}{dzd\Omega} dz,
\end{equation}
where $\Delta z$ is the width of the redshift bin.

\bibliographystyle{JHEP}
\bibliography{mybib}

\end{document}